\documentclass[a4paper,fleqn,usenatbib]{mnras}

\usepackage{newtxtext,newtxmath}
\usepackage[T1]{fontenc}
\usepackage{ae,aecompl}

\usepackage{graphicx}	% Including figure files
\usepackage{amsmath}	% Advanced maths commands
\usepackage{amssymb}	% Extra maths symbols
\usepackage{xcolor}
\usepackage[normalem]{ulem}
\usepackage{float}
\newcommand{\BH}{\textsc{BHrun}}
\newcommand{\NS}{\textsc{NSrun}}

\newcommand{\mdottot}{\dot{M}_\mathrm{tot}}
\newcommand{\mdotin}{\dot{M}_\mathrm{in}}
\newcommand{\mdotout}{\dot{M}_\mathrm{out}}

\newcommand{\DA}[1]{#1}

\title[Accretion onto NSs]{Radiative GRMHD simulations of accretion and outflow in non-magnetized neutron stars and ultraluminous X-ray sources}

\author[Abarca et al.]{
David Abarca,$^{1}$\thanks{E-mail: dabarca@camk.edu.pl (DA)}
W\l odek Klu\' zniak,$^{1}$
and Aleksander S\k{a}dowski$^{2}$
\\
$^{1}$Nicolaus Copernicas Astronimcal Center, Warsaw, Poland\\
$^{2}$Akuna Captial, 585 Massachusetts Avenue, Cambridge, MA 02139 
}

\date{Accepted XXX. Received YYY; in original form ZZZ}

\pubyear{2017}

\begin{document}
\label{firstpage}
\pagerange{\pageref{firstpage}--\pageref{lastpage}}
\maketitle

\begin{abstract}
We run two GRRMHD simulations of super-Eddington accretion disks around 
a black hole and a  \DA{non-magnetized, non-rotating} neutron star. 
\DA{The neutron star was modeled using a reflective inner boundary condition}. We observe the formation of a transition layer in the
\DA{inner region of the disk in the}
neutron star simulation which leads to a larger mass outflow rate and a lower radiative luminosity over the black hole case. Sphereization of the flow leads to
an observable luminosity at infinity around the Eddington value
\DA{when viewed from all directions} for the neutron star case, contrasting
to the black hole case where collimation of the emission leads to observable luminosities 
about an order of magnitude higher \DA{when observed along the disk axis}.
We find the outflow to be optically thick to scattering,
which would lead to the obscuring of any neutron star pulsations observed in corresponding %non-magnetized  
ULXs.  
\end{abstract}


%%%%%%%%%%%%%%%%%%%%%%%%%%%%%%%%%%%%%%%%%%%%%%%%%%

%%%%%%%%%%%%%%%%% BODY OF PAPER %%%%%%%%%%%%%%%%%%

\section{Introduction}

The transfer of matter with angular momentum 
onto a compact object \DA{occurs via} an accretion disk through which mass
and angular momentum are transported in opposite directions though
viscous processes \citep{ss+73}. The viscous process is now believed to be
magnetic turbulence induced by the magnetorotational instability \citep{balbus+91}.

One particular class of accreting objects 
which has gained interest in recent years are ultraluminous
X-ray sources (ULXs). These are X-ray bright objects observed 
\DA{outside the centers of galaxies} 
with luminosities 
\DA{from $10^{39} \sim 10^{41}$} ergs s$^{-1}$. 
\DA{The first explanations for such bright X-ray objects favored the elusive intermediate mass black 
holes radiating at or below the Eddington luminosity \citep{colbert+99}.}
\DA{Up until the discovery of ULXs, the most luminous 
	stellar-mass, persistent X-ray source was known to be Sco X-1, a 
	neutron star radiating at its Eddington limit at around $10^{38}$ erg s$^{-1}$ \citep{shklovsky+67,bradshaw+99}.} 
Currently, the leading explanation for ULXs is beamed emission from 
accretion in an X-ray binary
\citep{king+01}, implying that or super-Eddington accretion is responsible for the large observed luminosities. 
In particular, a  set of three such  
objects were observed which reveal X-ray pulsations 
with a period on the order of one second \citep{bachetti+14,furst+16,israel+17a,israel+17b}
\DA{excluding black holes as the accreting objects in these three sources}.
It can now be said with some certainty that a large fraction of 
ULXs are accreting neutron stars \citep{kluzniak+15,king+17,wiktorowicz+17,pintore+17}. 
\DA{There have been a wide range
of proposed values for the strength of the magnetic field from 
relatively low ($B<10^9$ G) \citep{kluzniak+15}, to moderate
($10^{10} \text{ G} \lesssim B \lesssim 10^{13}$ G) \citep{king+17, walton+18}, 
to high ($B>10^{13}$ G), magnetar-like fields
\citep{eksi+15,mushtukov+15b}.}

Accretion onto a neutron star (NS) is more complicated than accretion onto a 
black hole. Neutron stars have no event horizon. They
have a surface layer and outer crust which can reach densities of up to 
$~10^{11} \,\text{g/cm}^3$ \DA{at its base, where it is composed of fully ionized neutron rich nuclei.
At larger radii, and lower densities ($\sim 10^7$ g cm$^{-3}$) the composition of
the nuclei becomes less neutron rich.
Below $10^4$ g cm$^{-3}$, the nuclei are no longer fully ionized.
Finally, near the surface, the outer crust is composed mainly of crystalized iron atoms 
reaching down to $10$ g cm$^{-3}$ \citep{chamel+08}.}

%\pagebreak %% to fix pdflink problem
Gas accreting onto these outer layers is 
expected to slow down and release some of its kinetic energy 
\citep{ss+86, kluzniak+91, narayan+95, narayan+97,inogamov+99,sibgatullin+00, popham+01,mukhopadhyay+02}
 \DA{as it spins up the star \citep{kluzniak+85}}. 
This energy can be
converted into radiation (\DA{normally X-rays}) or transferred to the outflowing gas. 
Additionally, many neutron stars have strong magnetic fields, some of which can be
strong enough to channel the accreting gas into dense accretion columns, 
depositing gas at the magnetic poles, \DA{forming hot spots at low accretion rates, $\lesssim 10^{17}$ g s$^{-1}$}.
\DA{Misalignment of the magnetic poles with the rotation axis causes the hot spots to rotate
resulting observationally in X-ray (accretion powered) pulsars.} 
At high accretion rates ($10^{17} \sim 10^{19}$ g s$^{-1}$), the gas is expected to experience a radiation shock 
and to form an accretion column above the neutron star surface through which gas sinks slowly through a dense radiation field to eventually settle on the NS surface \citep{basko+76,mushtukov+15a,mushtukov+15b,revnivtsev+15}. 
\DA{As the accretion rate increases, the accretion column starts to widen and
 spread over a surface roughly corresponding to the
	surface of the magnetosphere. 
	Emission from the central regions and through the sides of the inner part of the accretion column can interact
	with the outer parts of the accretion column producing complicated pulse profiles 
	\citep{mushtukov+18}.}

\DA{At even higher accretion rates ($\gtrsim 10^{19}$ g s$^{-1}$), 
in the context of ULXs, the accretion column spreads into an accretion curtain,
a geometrically extended surface 
corresponding to an optically thick layer completely surrounding the pulsar 
magnetosphere which reprocesses all of the radiation generated near the
neutron star surface strongly smoothing the pulse profile \citep{mushtukov+17}.
Observationally, in 
pulsating and non-pulsating sources,
 this manifests as a double black body with hot (>1kev) and cold (<0.7 kev)
components
corresponding to the thermal emission from the accretion enveloping the magnetosphere, and
thermal emission from the accretion disk truncated at the magnetosphere, respectively \citep{koliopanos+17}. }

\DA{Here, using numerical simulations, we try to see if non-pulsating ULXs 
can be explained by super-Eddington accretion onto neutron stars.}
We ignore the effects of a stellar magnetic field,
(for an accretion rate of ten times the Eddington limit and a magnetic moment of
$\mu<10^{27}$ G cm$^3$ a simple calculation of the Alfv\'en radius shows the effects of the magnetic field are confined close to the neutron star surface), and consider only the effects of a hard surface. 
We use a general relativistic radiation magnetohydrodynamic (GRRMHD) code 
\textsc{Koral} to capture the most relevant physical processes. 

\subsection{Neutron star related accretion simulations}

In this section we mention some simulations that are related to 
neutron star accretion. \DA{The X-ray spectra from spherical accretion onto
high and low mass neutron stars was computed from coupled hydrodynamic 
radiation transfer calculations,{\tiny } which were shown to yield results which 
differ strongly from a black body \citep{alme+73}.}
\citet{dhang+16} performed 
hydrodynamic simulations of spherical accretion onto a hard surface in 
one and two dimensions. The hard surface was modeled in two ways, with a
reflective, and a `leaky' boundary condition, the latter being where mass is 
allowed to cross the inner boundary at a fixed subsonic speed to model efficient
cooling. \DA{This is important because not all works include a hard surface}.
More complicated Bondi-Hoyle \citep{mellah+15}  
and magnetic Bondi-Hoyle \citep{toropina+12} simulations have been performed, but they are without a hard inner boundary.

A 1.5D coupled radiative transfer and hydrodynamics calculation
was performed by \citet{kluzniak+91} which simulated the boundary layer between
the neutron star and the accretion disk.
\DA{The boundary layer was simulated by introducing an
optically thin stream of plasma inside the innermost stable circular orbit 
(ISCO), where the infall velocity quickly becomes supersonic. The plasma  decelerates
in the upper layers of the boundary layer on
the neutron star surface resulting in the creation of hard X-rays. 
Because velocities in the accretion gap are supersonic, such a calculation is valid without
considering the contribution of the accretion disk.}

A further work of interest is \citet{kawashima+16} who
performed radiation hydrodynamic simulations of the accretion 
column of a super-Eddington accretion neutron star using flux-limited diffusion, 
where the radiative flux follows the gradient of radiative energy density.
They found sub-Eddington luminosities along the optically thick accretion column
but super-Eddington luminosity when viewed from the sides,
in agreement with \citet{basko+76}. 

\citet{romanova+12} have performed global MHD simulations of MRI-driven accretion
onto magnetized stars. A number of interesting results are presented on the interaction
between the stellar magnetic field and the accretion disk, however, the lack of strong
gravity or radiation hydrodynamics means that the results \DA{do not accurately describe accretion onto
neutron stars at large accretion rates}.

\DA{\citet{parfrey+17} used an innovative method to run GRMHD simulations of accretion onto 
rotating magnetized neutron stars to model accreting millisecond pulsars. Their method interpolates between
the normal GRMHD flow and the force-free magnetosphere. At the lowest magnetizations of the neutron star, they show
that the magnetic field is crushed by the accretion flow, and accretion proceeds normally. Due to the lack
of radiation, their simulations are scale-free. When scaling
their system to the mildly super-Eddington accretion flow that we describe in this work, 
we find that indeed the magnetic field would be crushed at a 
magnetic moment of $\mu=10^{26} \text{ G cm}^3$, even when rotating at millisecond periods.}

\citet{takahashi+17} have published the first global 2.5D GRRMHD simulations of accretion onto a neutron star and their work represents the state-of-the-art
on the subject. 
They simulate super-Eddington accretion onto a magnetized neutron star 
with \DA{the radial flux and velocity set to zero at the inner boundary}, as a 
means to model a neutron star ULX system. They report luminosities of about an order of
magnitude above the Eddington limit with a significant amount of 
beaming which accurately describes a non-pulsating ULX source.
In their simulation, the magnetic field is strong enough to truncate the disk leading
to accretion along magnetic field lines.
They observe some matter piling up at the inner boundary due to the 
\DA{inner} boundary condition, but do not run the simulation for long duration, $(t_\mathrm{max} = 15,000 t_g)$,
and so it is hard to say what the effect of the accumulation of gas has on the accretion disk. 

Our work
considers the context of a boundary layer
(as opposed to an accretion column) with an accretion disk, 
with a sophisticated radiation treatment which deals with the 
optically thick and optically thin regimes.
In our simulation, we will focus strictly on the effect of a reflective boundary. We will run our simulations
for longer durations, \DA{$t\sim 160\,000\,t_g$}, where $t_g = GM/c^3$, to see what happens when a large amount of gas is accumulated in the vicinity of the neutron star.

\begin{figure*}
	\centering
	
	\begin{tabular}{cc}
		\includegraphics[width=\columnwidth]{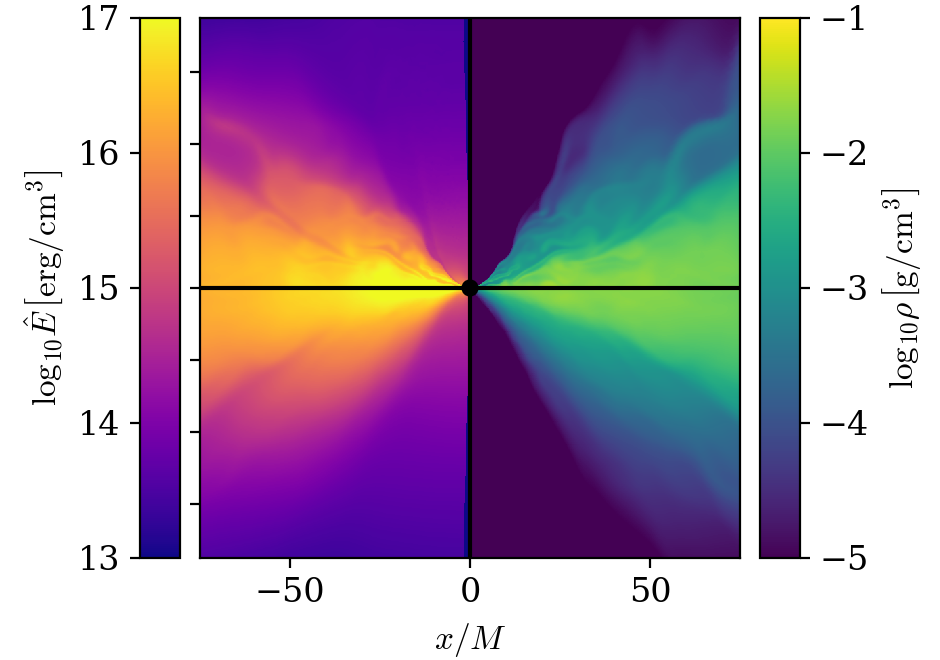} & 
		\includegraphics[width=\columnwidth]{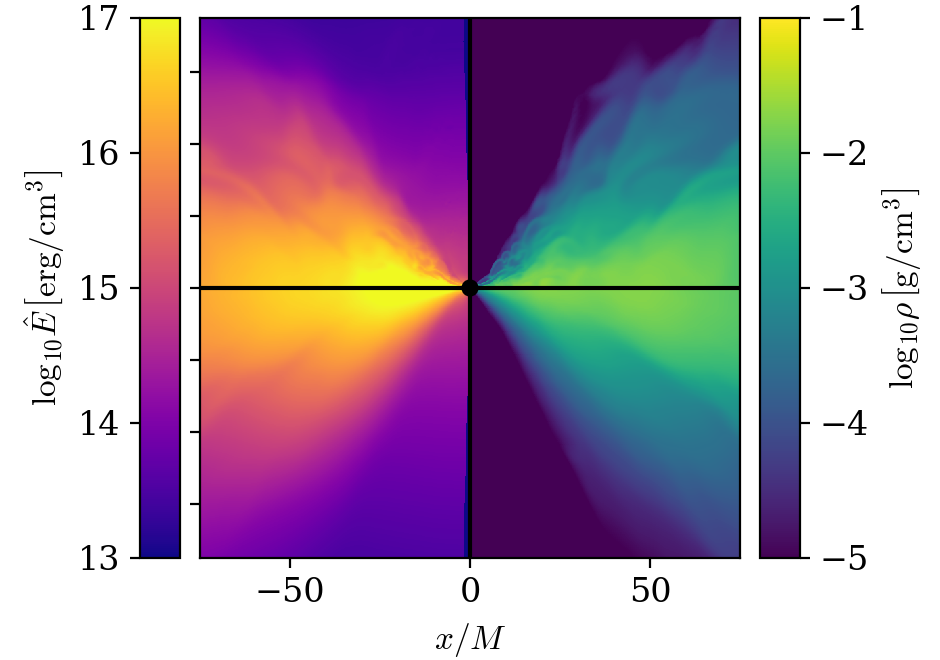} \\
		\includegraphics[width=\columnwidth]{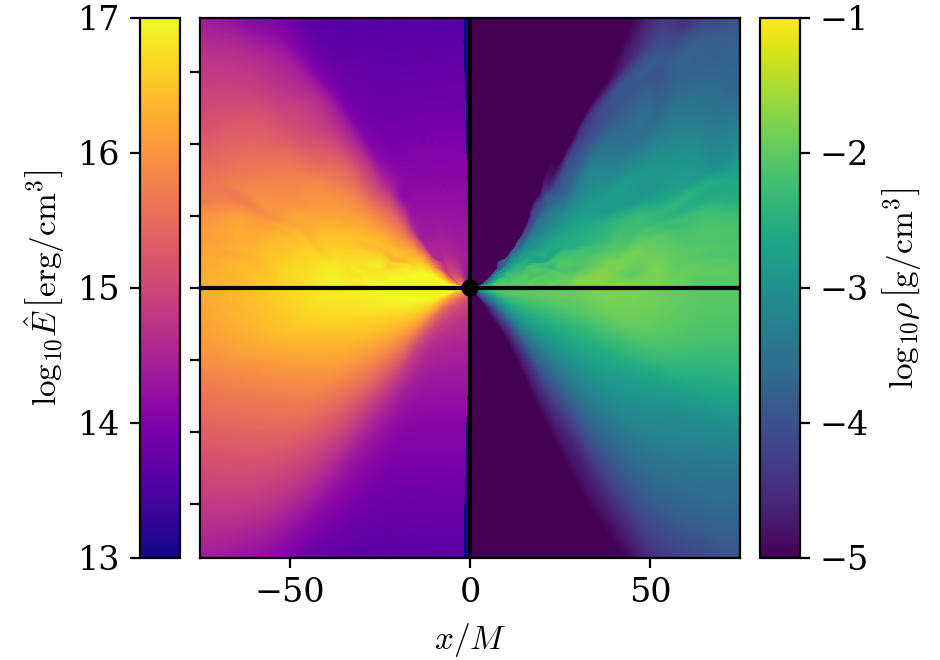} &
		\includegraphics[width=\columnwidth]{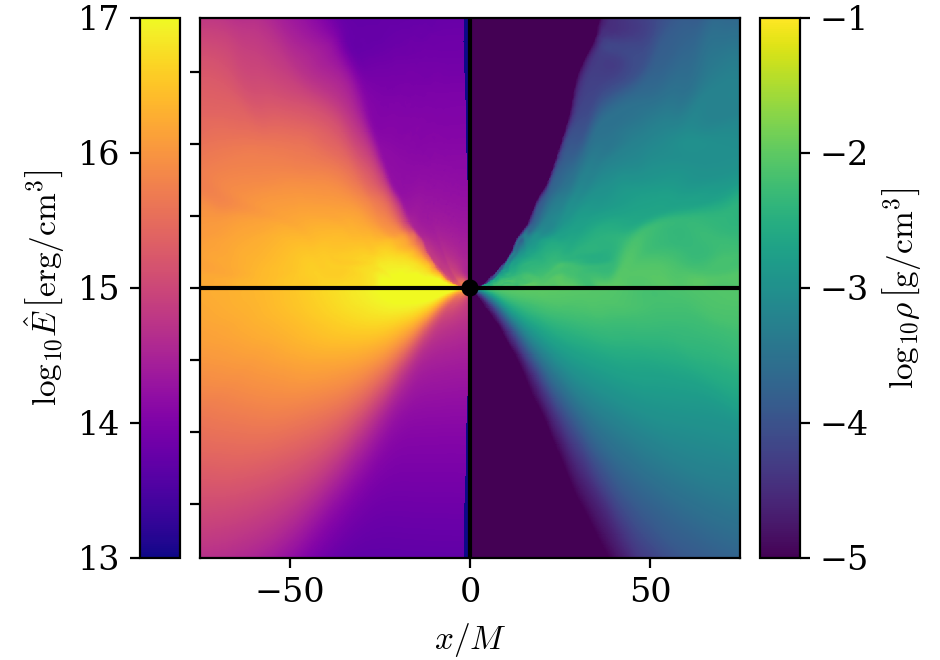}
	\end{tabular}
	\caption{Here we show snapshots of the 2.5D axisymmetric simulations of
		super-Eddington accretion from \BH. The left two
		quadrants of each image show radiative energy density in the fluid frame, and the right
		panels shows the rest mass density in the fluid frame. The upper panels of each image show 
		the instantaneous state of the simulation at time $t_i$, 
		and the bottom panels are time averaged from time $t_i-t_i/3$ to $t_i+t_i/3$. 
		From left to right and top to bottom the times correspond to, 
		$t_i  =  7500\, t_g, 15\, 000\, t_g, 30\, 000\, t_g$, and  $60\, 000\,t_g $. 
		\label{fig:bhframes}}
	
\end{figure*}

\begin{figure*}
	\centering
	
	\begin{tabular}{cc}
		\includegraphics[width=\columnwidth]{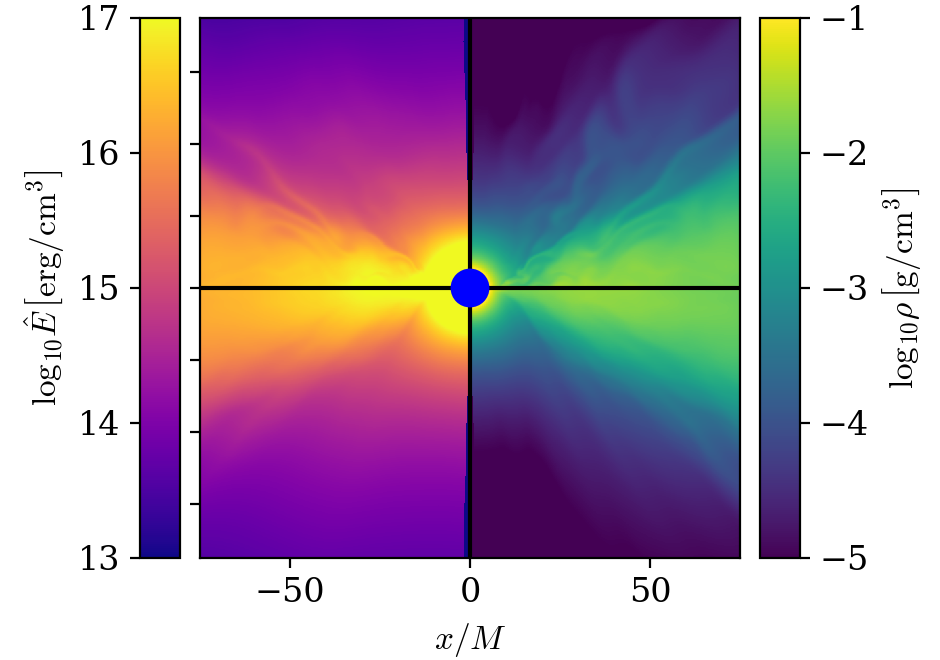} & 
		\includegraphics[width=\columnwidth]{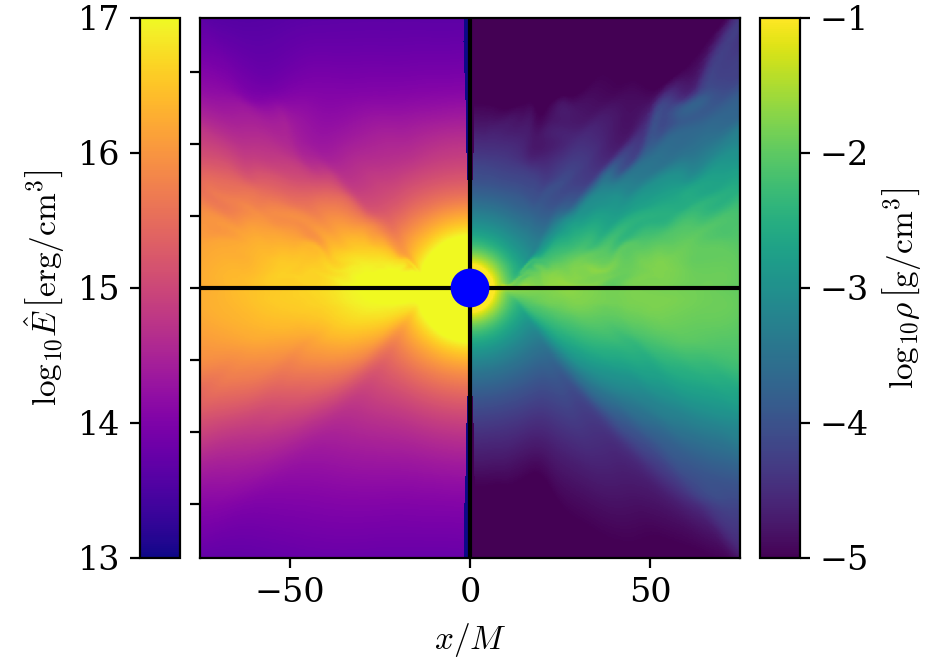} \\
		\includegraphics[width=\columnwidth]{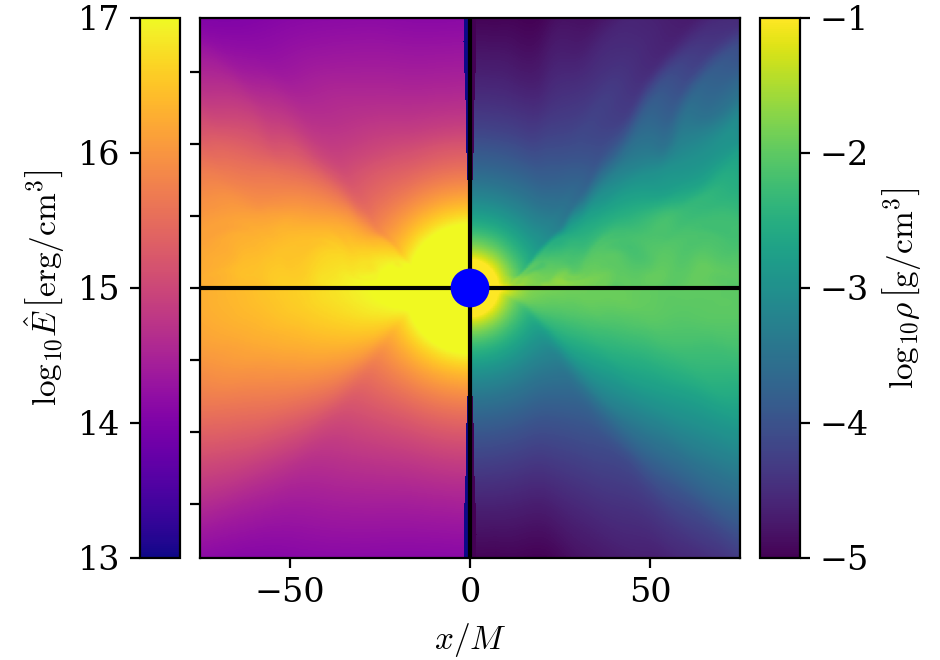} &
		\includegraphics[width=\columnwidth]{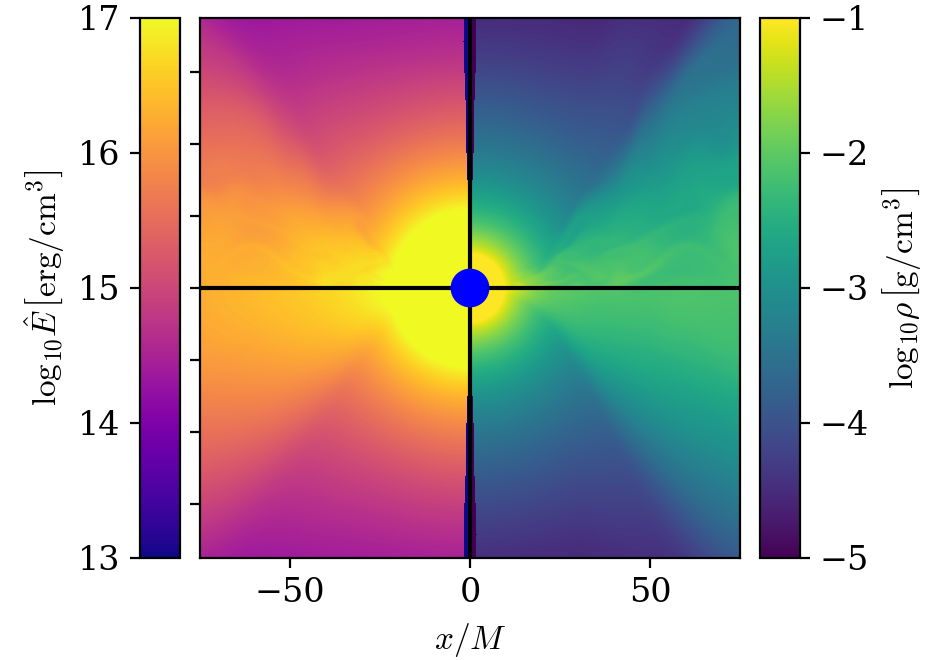}
	\end{tabular}
	\caption{Here, using the same scheme as in Fig.~\ref{fig:bhframes}, we show
		four snapshots of \NS, the simulation with the reflective boundary condition.
		\label{fig:nsframes}}
\end{figure*}

\section{Numerical Methods}
We investigate accretion onto neutron stars using a 
sophisticated 3D General Relativisic Radiation
Magnetohydrodynamics solver, \textsc{Koral} used extensively to study 
accretion onto black hole at high and low accretion rates, and other related 
phenomena. Details of the numerical implementation are
given in \citet{sadowski+m1,sadowski+dynamo}. Here we describe the most 
relevant features.

\subsection{Governing equations}
The equations of GRRMHD, which can be written in their conservative
form as
\begin{align}
&\nabla_\mu (\rho u^\mu) = 0, \\
&\nabla_\mu T^\mu\phantom{}_\nu = G_\nu, \\
&\nabla _\mu R^\mu\phantom{}_\nu = -G_\nu, \\
&\nabla_\mu(nu^\mu) = \dot{n},
\end{align}
are solved in \textsc{Koral} on a static one, two, or three dimensional mesh. 
Our mesh is a spherical 2.5D (2D axisymmetric) grid \DA{using a static metric, $g_{\mu \nu}$
with signaure ($-\, +\, +\, + $)}.
Here, $\rho$ is the gas rest-mass density in the comoving fluid frame, $u^\mu$ is the 
gas four-velocity, $T^\mu\phantom{}_\nu$ is the MHD stress-energy tensor 
given by
\begin{equation}
T^\mu\phantom{}_\nu = 
(\rho + u_{\mathrm{int}} + p + b^2)u^\mu u_\nu + 
\left(p+\dfrac{1}{2}b^2\right)\delta^\mu_\nu - b^\mu b_\nu. 
\end{equation}
Here, $u_{\mathrm{int}}$ is the internal energy of the gas, and
$p=(\gamma-1)u_{\mathrm{int}}$ is the gas pressure,  calculated using the 
adiabatic index, $\gamma=5/3$. The radiation stress-energy tensor is given by 
$R^\mu\phantom{}_\nu$ which is coupled to the gas stress-energy tensor by
the radiation four-force, $G_\nu$, making use of electron scattering and 
bremsstrahlung opacities as well as Comptonization 
\cite{sadowski+comp} which evolves the photon number, $n$, by taking into 
account the creation and annihilation of photons by emission and absorption, $\dot{n}$,
while conserving $n$ for Compton scattering exchanges of energy.
The radiation stress-energy tensor is completed using the $M_1$ closure scheme 
\citep{sadowski+m1}, which assumes there is a frame in which the 
radiation is isotropic. The $M_1$ scheme allows radiation to diffuse through
gas at large optical depths, and to freely stream along geodesics at very low
optical depths.
The magnetic field four-vector,
described by \citet{gammie+03}, is given by
$b^\mu$, and it is evolved using the induction equation which, when written 
in the coordinate basis appears as
\begin{equation}
\partial _t(\sqrt{-g}B^i) = -\partial_j\left(\sqrt{-g}b^j u^i - b^i u^j\right).
\end{equation}
Here $B^i$ is the normal magnetic field three-vector, which is related to the
magnetic field four-vector by
\begin{align}
b^t &= B^i u^\mu g_{i\mu}, \\
b^i &= \dfrac{B^i + b^t u^i}{u^t},
\end{align}
for metric $g_{ij}$ and metric determinant, $g$ \citep{komissarov+99}.

\subsection{Mean-field dynamo for 2.5D runs}
One particularly useful tool implemented in \textsc{Koral} is a mean-field 
magnetic dynamo which allows for axisymmetric 2D (2.5D) accretion disk simulations 
to be run for long durations without depleting the magnetic field due to 
turbulent dissipation, which normally occurs in axisymmetric
simulations of MRI \citep{sadowski+dynamo}. 
The dynamo has been tested against 3D 
simulations and has been found to accurately approximate the disk's spatial 
properties, accretion rate, surface density, and angular momentum for example.
The 2D dynamo disk does however have a tendency to overestimate the magnitude
and variability of the radiative flux \citep{sadowski+3d}. Nevertheless, the 
advantages of being able to run a 2D simulation as opposed to 3D make the
mean-field dynamo a valuable tool, allowing for almost a 100-fold speedup 
in runtime, and so we chose to implement it in this work in order to 
run long duration simulations. We also implement an adaptation to make
the dynamo more suitable to simulations where we expect a significant
amount of gas to accumulate at the inner simulation boundary. We
include a smooth cutoff to deactivate the dynamo in cells with
a specific angular momentum lower than $~$80\% of the Keplerian
value.

\section{Numerical Setup}

\subsection{Initial conditions }
We initialize our accretion disk in a typical way by starting with an equilibrium
torus near the non-rotating black hole as given in \citet{penna+13}. 
The torus is threaded with a weak magnetic field in loops 
of alternating polarity. The total pressure is distributed
between gas and radiation assuming local thermal equilibrium. Once the simulation
starts, the MRI quickly develops turbulence and accretion begins. We measure the rate
of mass accretion in units of 
$L_\mathrm{Edd}/ c^2$, where $L_\mathrm{Edd}$ is the
Eddington luminosity and 
$c$ is the speed of light. The initial torus is set up to give a constant accretion rate
of about 200 $L_\mathrm{Edd}/c^2$ which would correspond to
a luminosity of about 10 $L_\mathrm{Edd}$ for the efficiency of a Shakura-Sunyaev disk.

\subsection{Boundary conditions}
A common practice when simulating accretion onto stars is to ignore the
effects of a hard surface and let gas flow through the inner boundary in a
standard outflow boundary condition for the hydrodynamic quantities. 
This allows the simulation to approach a quasi-steady
state and is useful for studying the interaction between the stellar magnetic field, and 
the accretion disk \citep{romanova+12,cemeljic+13}. This allows the star to behave 
somewhat like a black hole.\footnote{This is not entirely true, there are a variety of inner 
	boundary conditions on the magnetic field quantities that have various effects on the
	absorption of the hydrodynamic quantities.} In order to study the the difference between an
inflowing boundary condition at the inner edge, and reflective boundary condition, as well
as to have a natural, physical inner boundary as a baseline comparison, we run one simulation
with a black hole as the inner boundary condition. This is achieved by choosing a horizon
penetrating coordinate system (Kerr-Schild) and placing the inner boundary of the simulation
sufficiently behind the event horizon. This simulation we call \BH{}.

In order to study the effects of the release of kinetic energy, which is expected to 
significantly impact the behavior of the accretion disk and outflows, we implement a 
reflective boundary for \DA{the main simulation of our study}. 
The reflective boundary at $r=r_\mathrm{in}$ is set up so that the reconstructed radial velocity of the gas
at the inner boundary is \DA{opposite about} zero \DA{($u^r_l = -u^r_r$ where $u^r_l,u^r_r$ are the left and right reconstructed radial velocities at the cell interface of the inner boundary)}, so that no gas is able to leave the domain. 
We also set the tangential velocities $u^\theta, u^\phi$ in the ghost cells to zero. 
Note that this does not enforce the reconstructed tangential velocities to be exactly
zero at the inner boundary, it does however effectively remove angular momentum and allows
the gas to approach a non-rotating state at the inner boundary, and so we expect the formation
of a boundary layer. Radiative quantities are treated in the same way. We expect these 
boundary conditions to more accurately reflect the behavior of an accretion disk 
and a boundary layer around a neutron
star than an inflow boundary condition, and so this simulation we call \NS{}.

\subsection{Grid}
\DA{
Both simulations are run in Kerr-Schild coordinates 
(although this is not strictly necessary for \NS{}) 
with a non-rotating, central mass of $M=1.4M_\odot$ (Schwarzschild spacetime).
We run two simulations on \sout{identical} 2D, axisymmetric spherical  grids with logarithmic spacing in radius, and increased resolution near the equatorial plane.  The resolution for \NS{}
is $N_r \times N_\theta\times N_\phi = 352 \times 240 \times 1$. The radial coordinate 
spans from $r_\mathrm{in} = 5 r_g$ to $r_\mathrm{out} = 5000 r_g$ where $r_g = GM/c^2$.  
The resolution was higher for \BH{} ($N_r \times N_\theta\times N_\phi = 384 \times 240 \times 1$) because we have
to extend the inner boundary to below the event horizon at $r=2 r_g$. 
The larger number of radial grid cells allows the two
simulations to have comparable resolution at the same radii. }

\section{Results}

We have run two axisymmetric GRRMHD simulations, one with a reflective, non-rotating inner boundary at radius $r=5r_g$ \NS{},
and one with a black hole inner boundary, \BH{}.
Snapshots and time averages of both simulations are shown in Figs.~\ref{fig:bhframes}~\&~\ref{fig:nsframes}. 
Each figure contains four images, each corresponding to a different time, 
$t_i  \in \left\{ 7500\,t_g, 15\,000\,t_g, 30\,000\,t_g, 60\,000\,t_g \right\} $. Snapshots at these
times are shown in the upper two quadrants of each image. We are also showing the 
time averaged structure of the simulations in the lower two quadrants of each image, all 
of which are averaged from $t=t_i-t_i/3$ to $t=t_i+t_i/3$. The right two quadrants of each image
show rest mass density, and the left two quadrants show radiation energy density in the 
fluid frame.

\begin{figure*}
	\centering
	\begin{tabular}{cc}
	$\pi/2 \pm \pi/20$ & $\pi/8 \pm \pi/20$ \\
	\includegraphics[width=\columnwidth]{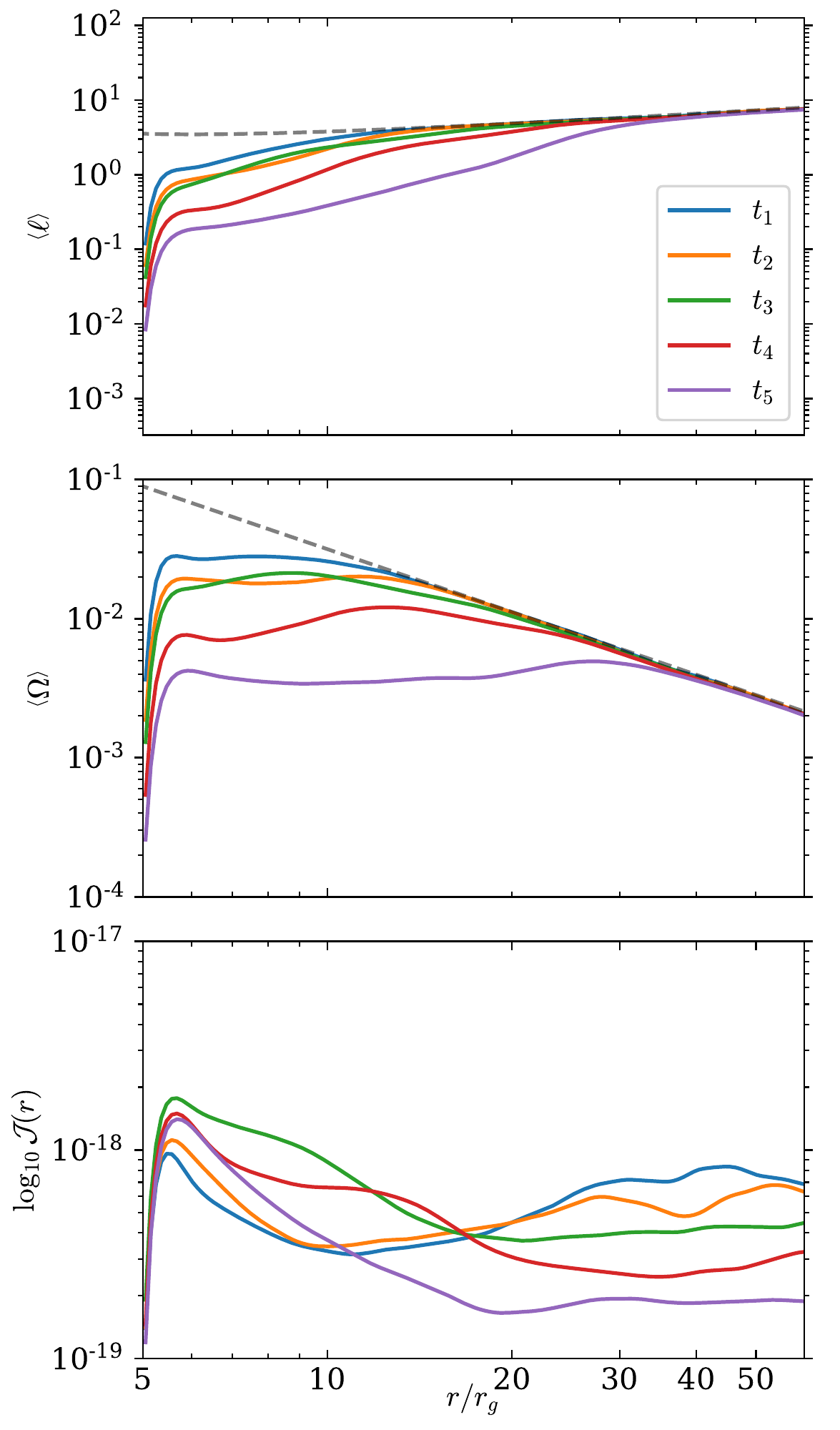} &
	\includegraphics[width=\columnwidth]{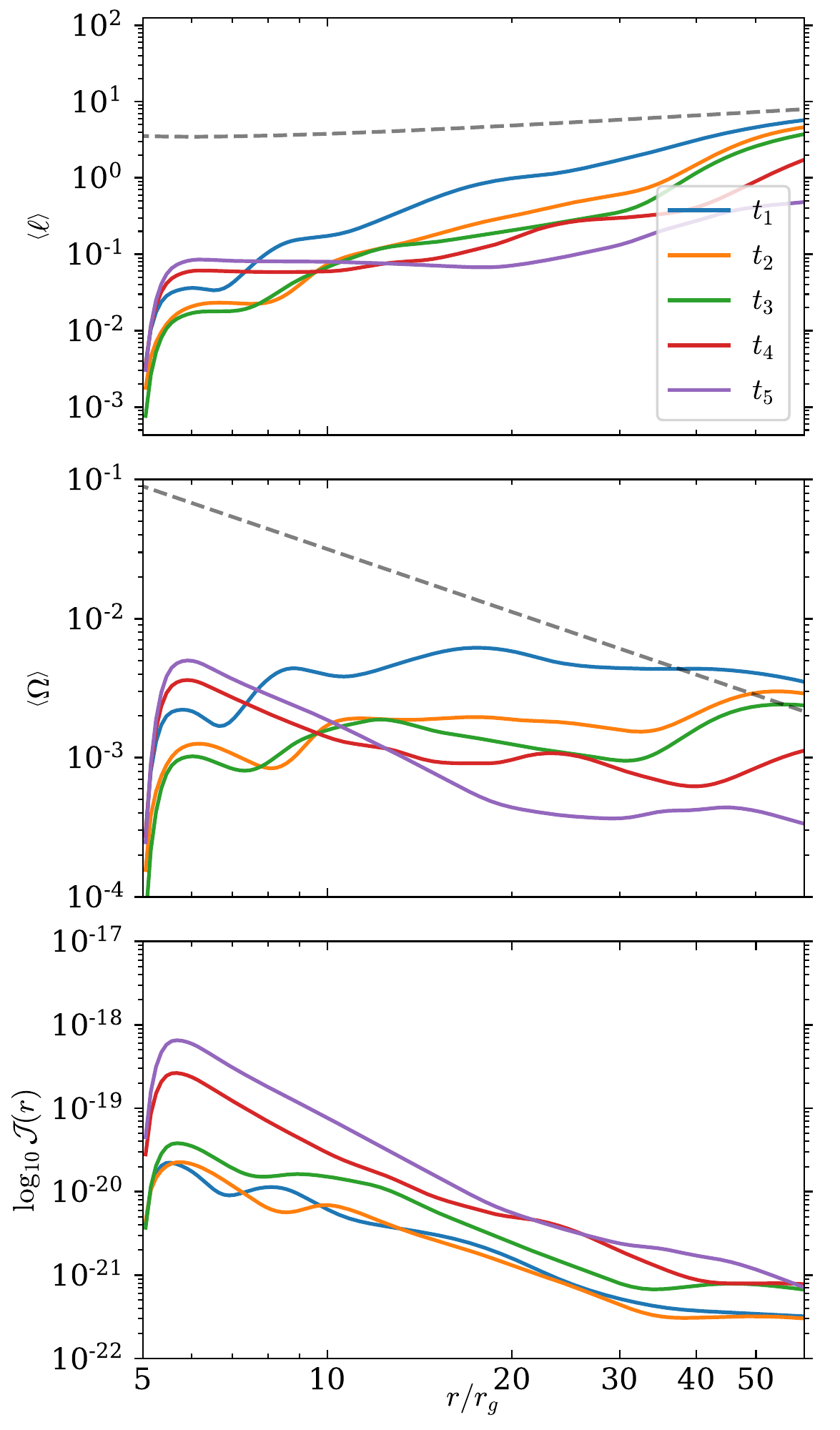} 
	\end{tabular}

	\caption{\DA{We plot angular momentum related quantities in the
		disk mid-plane ($\theta = \pi/2$, left) 
		and in the polar region of the simulation 
		($\theta=\pi/8$, right). The plots were averaged/integrated over $\theta$
		over a range of angles corresponding to $\pi/10$ radians.}
		Top: Here we show a time series of 
	the specific angular momentum of the gas, $\ell=u_\phi$ for \NS{}.
	The  plotted quantities correspond
	to a 
	$\rho$-weighted $\theta$-average of the time averaged simulation data.   
	The durations for the averaging are the same as for Fig.~\ref{fig:bhframes}~\&~\ref{fig:nsframes}, $t_1 = 7500\, t_g, t_2=15\, 000\, t_g, t_3=30\, 000\, t_g, t_4=60\, 000\, t_g$ \DA{with an additional time interval centered at, $t_5=120\, 000\,t_g$}. The Keplerian angular momentum is shown
	by the dashed grey line. 
	Middle: Here is the same plot as above except for angular velocity, $\Omega=u^\phi/u^t$. 
	Bottom: Here, we are plotting the total angular momentum density integrated along the $\theta$ direction.
		\label{fig:levo}}
		\end{figure*}

The \BH{} simulation forms a typical, geometrically thick, super-Eddington accretion 
disk similar to what is seen in previous simulations of this type,
 see \citet{yang+14,jiang+14,sadowski+dynamo, sadowski+3d,ogawa+17} for examples.
As the disk reaches inflow equilibrium one can see the formation of a funnel
region along the polar direction. The funnel region is nearly devoid of gas 
but is filled with radiation and so one can expect the disk to appear very bright 
when viewed along the funnel.

The reflective inner boundary shows a much different scenario. Gas cannot 
pass through the inner boundary, so it accumulates a dense layer around the inner 
edge of the simulation.
At late times, the density at the inner edge reaches nearly 10 g/cm$^3$. The density at the
inner edge increases gradually as more and more gas is compressed into the atmosphere.
We can also see large amounts of radiation being accumulated at the inner boundary due 
to a combination of dissipation and advection. 
In general, hot flows of this type are optically thick to scattering, but optically thin to absorption. However, it is unclear a priori if the
radiation can escape to infinity in any reasonable amount of time because gas is blown off the outer edges of the accumulating layer which forms a dense outflow. Photons produced near the 
disk must random walk through this thick outflow, some of them may even be scattered 
back through the disk. We are particularly interested in whether or not pulsations may be
visible. With such a thick scattering dominated atmosphere, it is likely that pulsations would
lose their coherence. We measure the scattering optical depth in the next subsection.

\subsection{Transition layer and accretion disk structure}
The inner boundary of the neutron star simulation 
causes the angular velocity of the gas to approach zero.
This is seen in Fig.~\ref{fig:levo}.  The top panels of Fig.~\ref{fig:levo} show the 
density-weighted, $\theta$-averaged profiles of the specific angular 
momentum of time averaged time intervals used in Figs~\ref{fig:bhframes} \&~\ref{fig:nsframes} for \NS{}.
\DA{Because \NS{} is computationally cheaper compared to \BH{} and because the behavior at late times 
is interesting we run \NS{} for an additional $80\,000$ $t_g$ so that we can show an additional time interval
in Fig.~\ref{fig:levo}, $t_5 \sim 120\,000\,t_g$}.
\DA{The left three panels show the midplane quantities, averaged or integrated
	over a range of $\theta$ from $\pi/2-\pi/20$ to $\pi/2+\pi/20$. The right
	three panels are integrated in the polar region, over $\theta$ from 
	$\pi/8-\pi/20$ to $\pi/8+\pi/20$.} 
As the simulation evolves, we can see the formation of a transitional region  which
expands outwards as more gas is accumulated. \DA{The rate of expansion measured by calculating the radius at which
$\ell$ reaches 90 percent of its Keplerian value follows a $t^{0.85}$ power law.}
\DA{For accretion onto neutron stars at lower accretion rates, this transitional layer usually 
lies in a small belt around the star where the flow properties transition to match the 
stellar surface. This is the classical picture of a boundary layer. We can see from Fig.~\ref{fig:nsframes}
that we have a much more extended, atmospheric layer, although the flow velocity does transition
to match the surface of the neutron star, so we will refer to this region as the transition layer.}

The process by which angular momentum is transferred from the accretion disk to a neutron star
is complicated.\DA{Various viscous and magnetic processes are involved. 
We try to reproduce the effect of driving the tangential
velocity to zero at the inner edge of the simulation with our inner boundary condition.
The normal magnetoturbulent processes that transport angular momentum through 
the disk do not operate on the numerical level of the cell interface. 
Instead angular momentum is transported between cells 
according to the flux computed by the HLL Riemann solver.
The angular momentum flux is made up of a hydrodynamic and a magnetic component,
 $T_{r\phi} = T^\mathrm{(HD)}_{r\phi} + T^{\mathrm{(mag)}}_{r\phi}$ where 
 $T^\mathrm{(HD)}_{r\phi} = (\rho + u_\mathrm{int} + p) u_r u_\phi$ and
 $T^\mathrm{(mag)}_{r\phi} = b^2 u_r u_\phi - b_r b_\phi $. By far the largest contribution
 at the inner edge is the $\rho u_r u_\phi$ term. The inner boundary condition leads to a flux
 that is appoximately, $T_{r\phi} \approx \rho u_r u_\phi /2$. Reflecting the angular velocity,
 for example, would lead to $T_{r\phi} \approx \rho u_r u_\phi$.
 We can see that the source of the torque at the inner edge is numerical in nature.
 A more detailed study of the effect of different boundary conditions on the transport of
 angular momentum through the transition layer is left to further studies. For now, we are
 satisfied the angular momentum transitions towards zero at the inner boundary
 through the midplane and along the disk axis.}

The middle two frames of Fig.~\ref{fig:levo} show angular velocity averaged in the same way
as the top panels of Fig.~\ref{fig:levo}. We
can see the expansion of the transition layer evolving into a quasi-flat region
in angular velocity in the midplane indicating that some large scale coupling
\DA{is causing the inner disk to rotate like a rigid body}.
The innermost region is driven
towards zero angular velocity as is expected by the boundary condition.

\DA{The polar region shows different behavior. The angular momentum in the 
	polar regions increases with radius at early times, and evolves to be 
	quasi flat at late times in the inner region. Meanwhile, the angular velocity
	starts quasi flat, and evolves to decrease with radius at late times, so the
	transition layer does not display rigid body behavior near the poles.}
	
As is seen in Fig.~\ref{fig:nsframes}, the transition layer reaches
very high densities, and so, even though it has very low angular velocity, it is still
able to contain a significant amount of angular momentum, as shown in the bottom panels 
of Fig.~\ref{fig:levo}. Here we are plotting the angular momentum integrated at a particular
radius, $r$, $\mathcal{J}(r) = 2\pi \int \rho u_\phi \sqrt{-g} d\theta$. The total angular momentum 
in the transition layer increases with time, especially near the inner edge.  
Angular momentum is transported through the transition layer where it accumulates at the
inner edge. It is likely that a stronger numerical torque at the inner edge would
allow more angular momentum to flow through the inner boundary, thus smoothing
$\mathcal{J}$ near the neutron star surface. Dynamically, what is important are $\ell$ and $\Omega$,
both of which are largely below their corresponding Keplerian values, and so we do not
expect a stronger torque at the inner boundary to have a significant effect on the 
evolution of the simulation. 

\begin{figure*}
	\centering
	\begin{tabular}{cc}
		\includegraphics[width=\columnwidth]{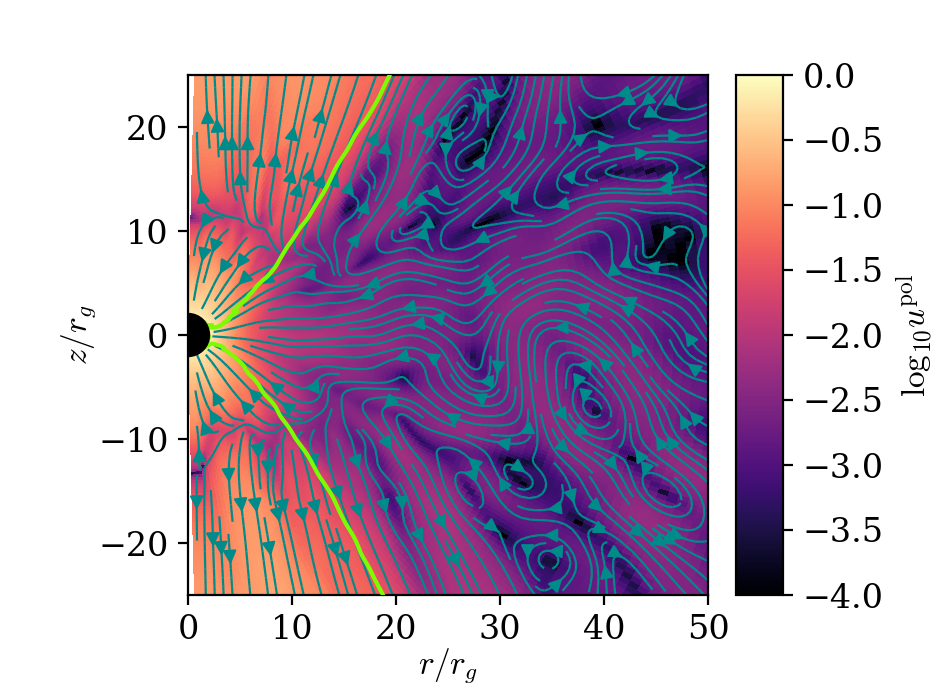} & 
		\includegraphics[width=\columnwidth]{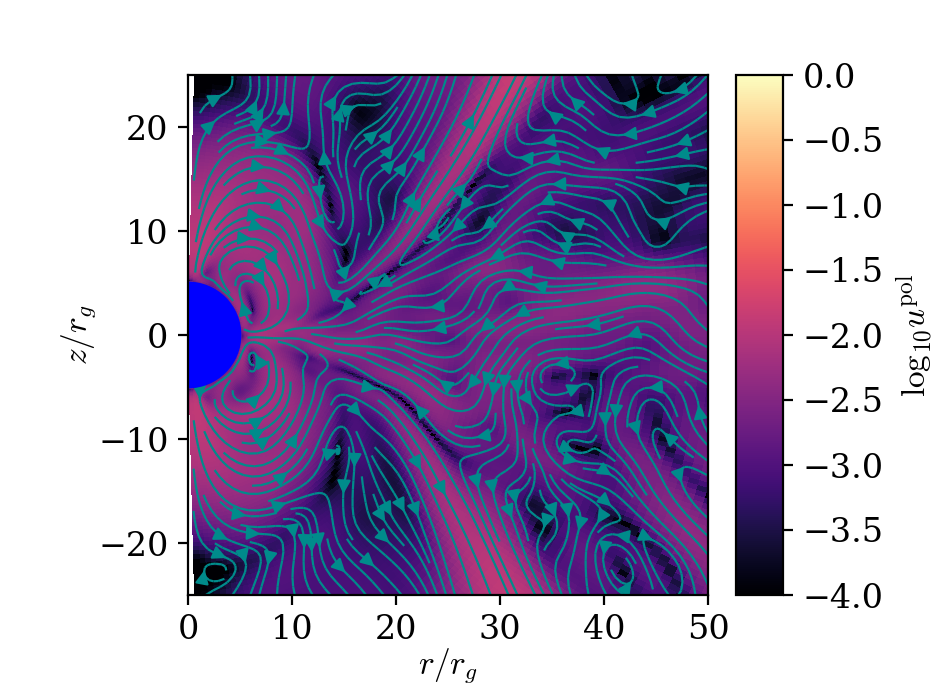} \\
		\includegraphics[width=\columnwidth]{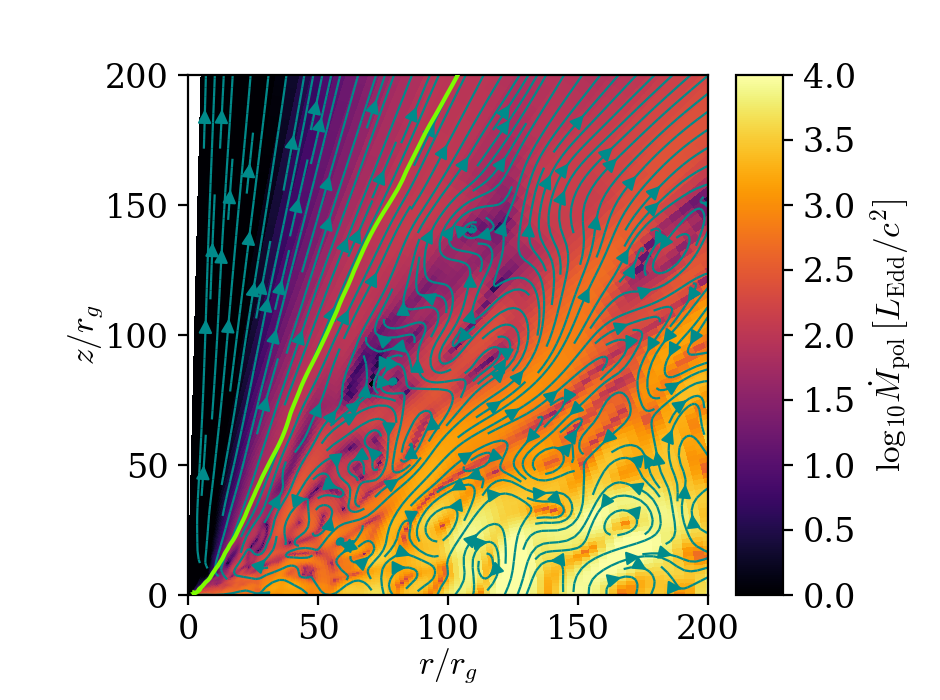} & 
		\includegraphics[width=\columnwidth]{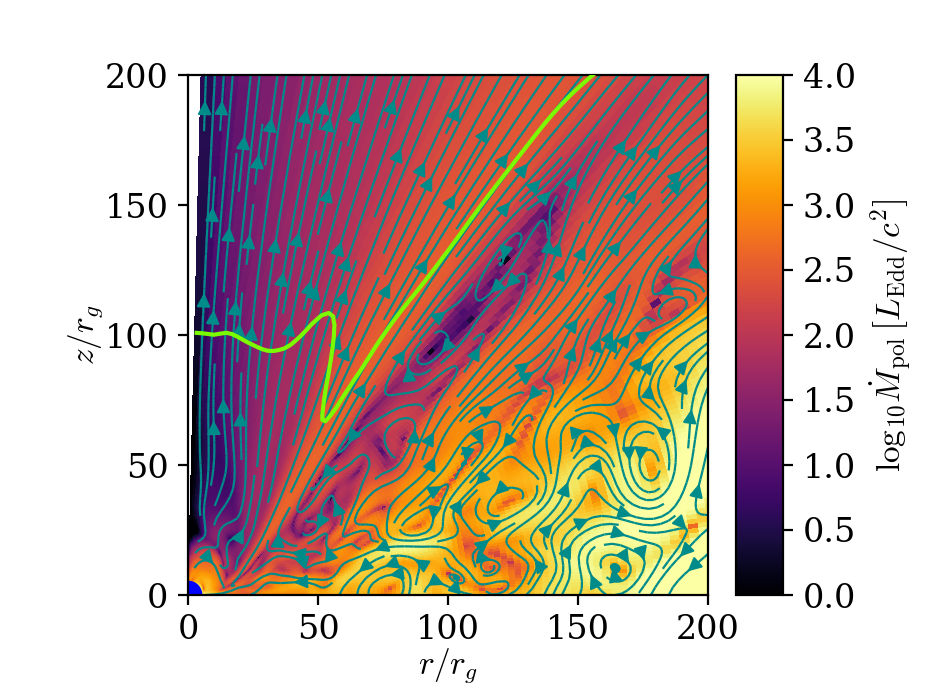} \\
		\includegraphics[width=\columnwidth]{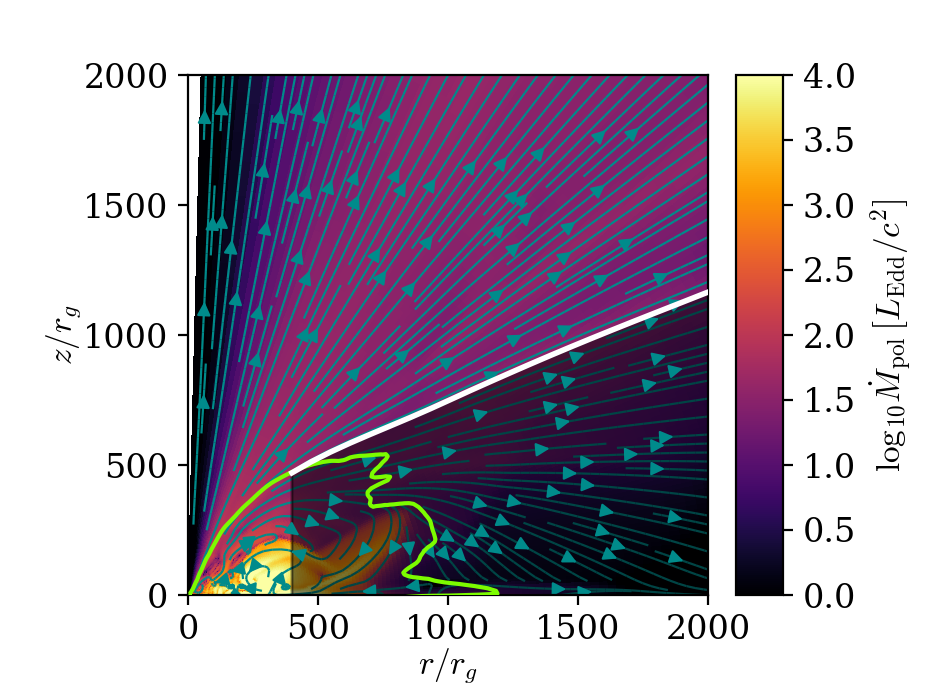} & 
		\includegraphics[width=\columnwidth]{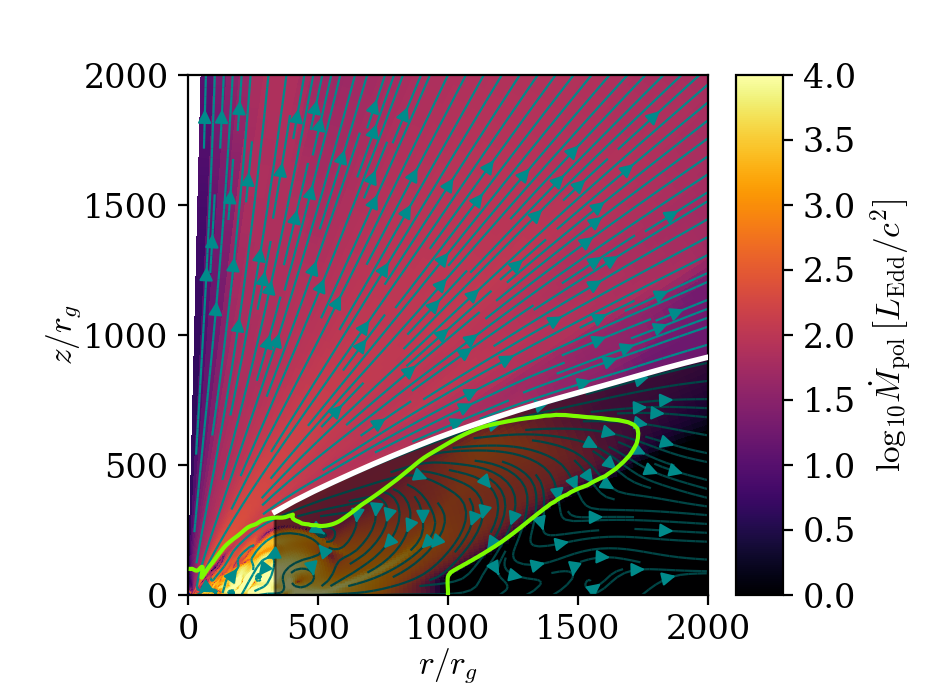} \\
	\end{tabular}
	\caption{ Here we show the time averaged poloidal velocity flow structure of the black hole-like simulation (upper-left) and reflective simulation (upper-right). 
		The magnitude of
		the poloidal velocity \DA{($u^\mathrm{pol} = \sqrt{u^r u_r + u^\theta u_\theta}$)} is given by the color map. 
		The streamlines follow the poloidal 
		velocity vectors for all plots in the figure. 
		The middle two images show the mass accretion rate in the poloidal plane
		\DA{($\dot{M}_\mathrm{pol} = 2 \pi r^2 \sin \theta \, \rho u^\mathrm{pol}$)}
		 for \BH{} simulation (middle-left) and \NS{} (middle-right).
		\DA{The two bottom images are the same as for the middle images, except
			for a larger range of $r$ and $z$. We show a contour of $Be=0$ in green
			and in white we show the last streamline which originates from the
			$Be=0$ surface. The shaded region below,  we exclude from our analysis.}
		\label{fig:velavg}}
\end{figure*}

The transition layer does not display much structure in the spatial 
distribution of gas density and radiation energy density, except for a strong
radial gradient, even when the color scales are suitably adjusted. 
The flow structure is much more informative. 

In Fig.~\ref{fig:velavg}, we show
the time-averaged spatial distribution of the poloidal velocities in the two upper panels for
the time interval, $t_4 $from $40,000\, t_g$ to $80,000\, t_g$. 
\DA{All of the remaining figures are constructed with time averaged data over the same period.}
\DA{The poloidal velocity and accretion rate
	are defined as follows,
	$u^\mathrm{pol} = \sqrt{u^r u_r + u^\theta u_\theta}$, 
	$\dot{M}_\mathrm{pol} = 2 \pi r^2 \sin \theta \, \rho u^\mathrm{pol}$.}
For \BH{}, the flow is primarily directed inward in
the disk and at the inner boundary. This is a boundary condition imposed by 
the space-time.
 In the polar
region, one can see the transition from inflow to outflow at about radius of 
$r=10 r_g$. This is the stagnation radius, above which is a radiation driven
outflow. Velocities in the polar region are relativistic, while velocities in the 
disk are around a few hundredths of the speed of light except in the very inner
regions.

Again, \NS{} shows a much different inner structure. Gas flows
though the transition layer in the equatorial plane where it meets the reflective
boundary and is directed tangentially along the surface of the inner boundary until
the polar regions where it is again redirected. The whole process forms two large
eddies which seem to recycle the gas into the inner edge of the accretion flow, which 
could be responsible for the large scale coupling seen in Fig.~\ref{fig:levo}.
\DA{The eddies seem to be connected to a conical outflow. We can 
	see two streams of gas being launched from the
	two regions where the eddy circles back to the disk.}
It is important to note that this is the time average structure, the non-averaged flow
being much more turbulent. 
\DA{The eddies are indicative of convective cells. Indeed, a calculation
	of the Schwarzschild stability parameter shows that the transition layer
	should be unstable to convection. Whether convection is driving
	the eddy motion is more difficult to say due to the extra source of momentum
	from the accretion flow.}

The poloidal accretion rate is shown in the middle two panels of Fig.~\ref{fig:velavg} over a large range in $r$  and $z$.
The accretion rate in the transition layer is nearly an order of magnitude higher 
than in the disk, indicating that the gas is recycled many times in the inner
flow. For \BH{}, we see the typical picture of a nearly 
empty funnel region, indicating that most of the outflowing gas is ejected in a 
wind at larger radii. 

\DA{We expect a much different picture in the presence of strong magnetic field. 
For $\mu\gtrsim 10^{30} \text{ G cm}^3$, we expect
the accretion flow to be directed along magnetic field 
lines and deposited at the poles for a dipolar magnetic field
as was seen in \citet{takahashi+17}. The magnetic field would also
arrest any outflows from the inner most regions, so it is likely that an optically thin
funnel would also be able to form in this case, leading to much larger observed luminosities. }

\DA{We show the poloidal accretion rate over an even larger range of 
	$r$ and $z$ in the lower two panels of Fig.~\ref{fig:velavg}. We can 
	no longer see
	the structure of the conical outflows in \NS{} which was
	decollimated at larger radii. We also show a contour of the 
	relativistic Bernouilii number in green, $Be=0$, which we define
	and discus in more detail in Section~\ref{sec:outflow}. 
	We can clearly see that more material is launched in 
	\NS{} than in \BH{}. }
	
	\DA{
	The region behind the initial torus in our simulation is unphysical. 
	In a real physical system
	this region would be occupied by the gas flowing from the companion star. 
	We exclude this region in a somewhat arbitrary way. We pick the last streamline 
	which originates from the surface defined by $Be=0$ (shown in white) which does not turn back towards the
	disk midplane at larger radii. The gas inside this region is expected to flow back towards the
	disk or to the companion star, and so we exclude this region from all of our calculation.}

\begin{figure*}
	\centering
	\begin{tabular}{cc}
		\includegraphics[width=\columnwidth]{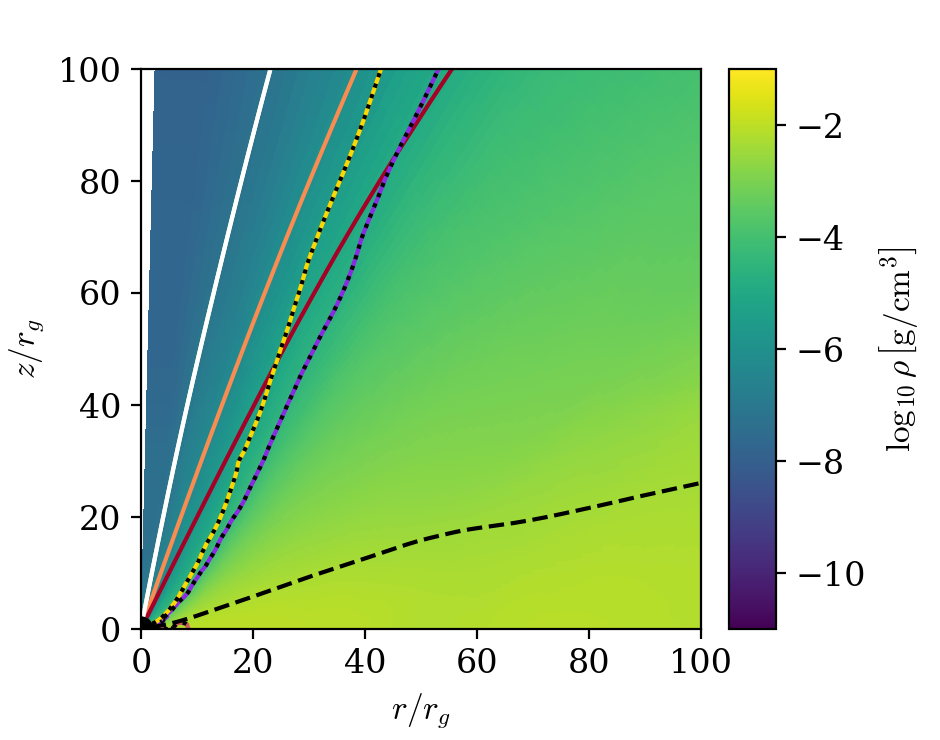} & \includegraphics[width=\columnwidth]{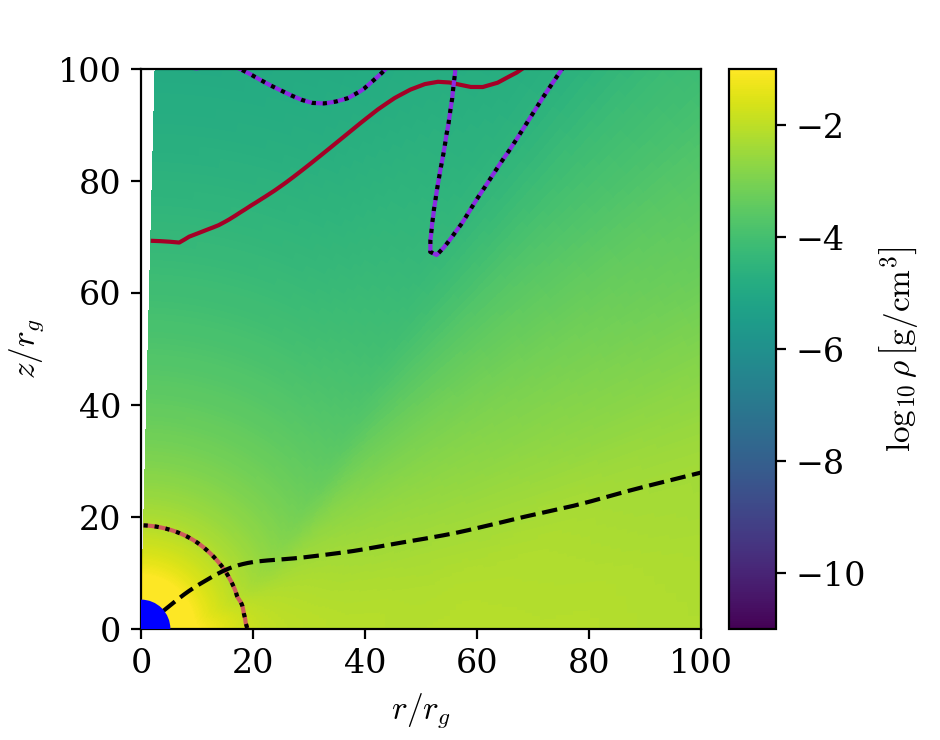}\\
		\includegraphics[width=\columnwidth]{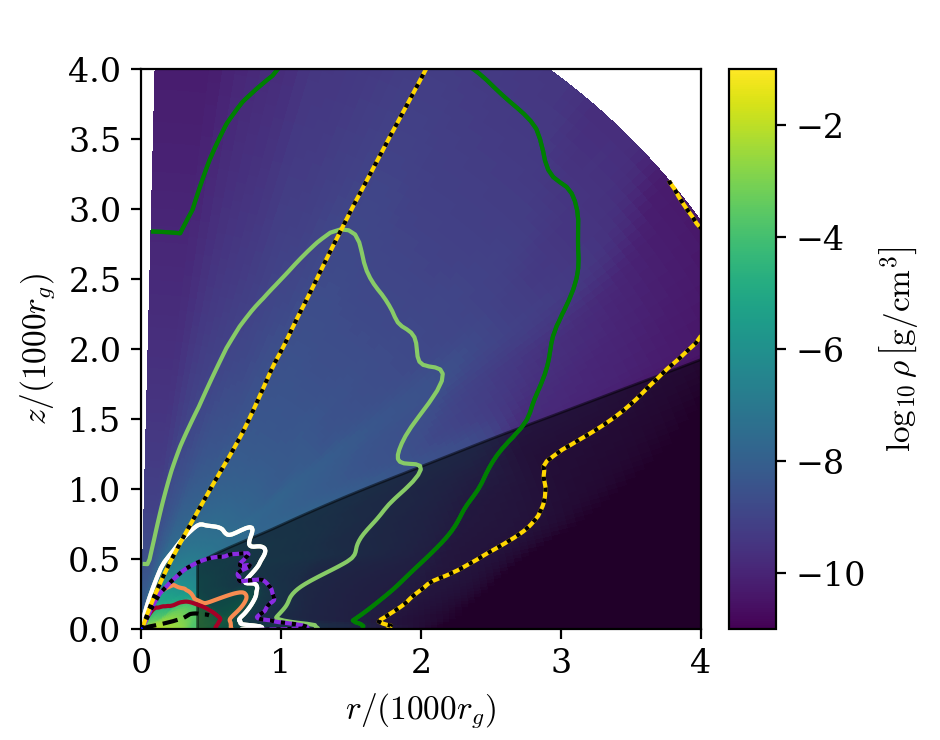} & \includegraphics[width=\columnwidth]{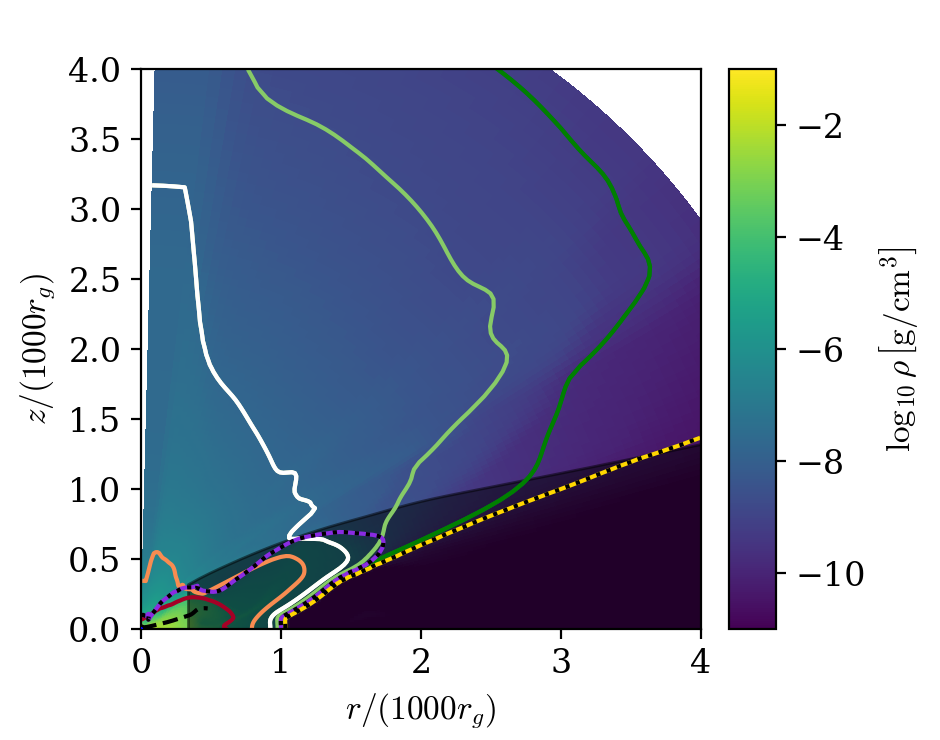} \\
	\end{tabular}
	\caption{Here we show the time averaged structure of the \BH{} (top-left) and the \NS{} (top-right). Expanded views
		of the simulations are shown in the bottom panels.
		The colormap shows density. The colored contours correspond to optical depth to scattering. From green
		to red the values correspond to $\tau_\mathrm{sca} = 0.01, 0.1, 1, 10, 100$. 
		The while line shows $\tau_\mathrm{sca}=1$.
		The striped red, purple, and yellow lines show contours of $Be=-0.05,0,0.05$
		respectively. \DA{The dashed black line shows the rms scale height, $h$}.
		\DA{We have also darkened the region contained by the last streamline to 
		originate from the $Be=0$ surface. }
		\label{fig:avg}}
	
\end{figure*}

In Fig.~\ref{fig:avg} we show the time averaged (again, over $t_4$) spatial structure of both disks. 
The colormap of density has been expanded to cover a wider range of densities with
the trade off being a lack of detail. We show two zoom levels. 
Red, white, and green contours of the 
scattering optical depth are also shown. From green to white to red, the contours correspond
to $\tau_\mathrm{sca} = 0.01, 0.1, 1, 10, 100$. 
\DA{Here, $\tau_\mathrm{sca}$ is calculated in a simple way,
\begin{equation}
\tau_\mathrm{sca}(r) = \int_r^{r_\mathrm{out}} \rho \kappa_\mathrm{es} \, \sqrt{g_{rr}} dr,
\end{equation}
where $\kappa_\mathrm{es} = 0.34$ g cm$^{-2}$}.
For reference, the
root-mean-squared polar scale height of the disk defined by 
$h = \sqrt{\int(\theta-\pi/2)^2 \rho d\theta/\int \rho d\theta}$.
For \BH{} (left), super-Eddington accretion leads to a thick disk with strong outflows and an optically thin funnel region reaching all the way down to the inner boundary. 
Again, we see a different picture for \NS{} (right). 
First, as is also seen in Fig.~\ref{fig:nsframes} a large amount of gas is deposited 
in a transition layer around the inner edge. In addition, a large amount of gas is
ejected and the entire domain is filled with a thick outflow. Measuring the optical 
depth shows that this outflow is extremely optically thick. The last
contour of the scattering optical depth visible in the figure is given by the red line which 
corresponds to $\tau_\mathrm{sca}$ of 100. A zoom of the entire simulation domain
is shown in the bottom two panels.
 We can see the photosphere for \NS{} lies on average at
a couple thousand $r_g$, which is indicated by the white line.

\begin{figure*}
	\centering
	\includegraphics[width=2\columnwidth]{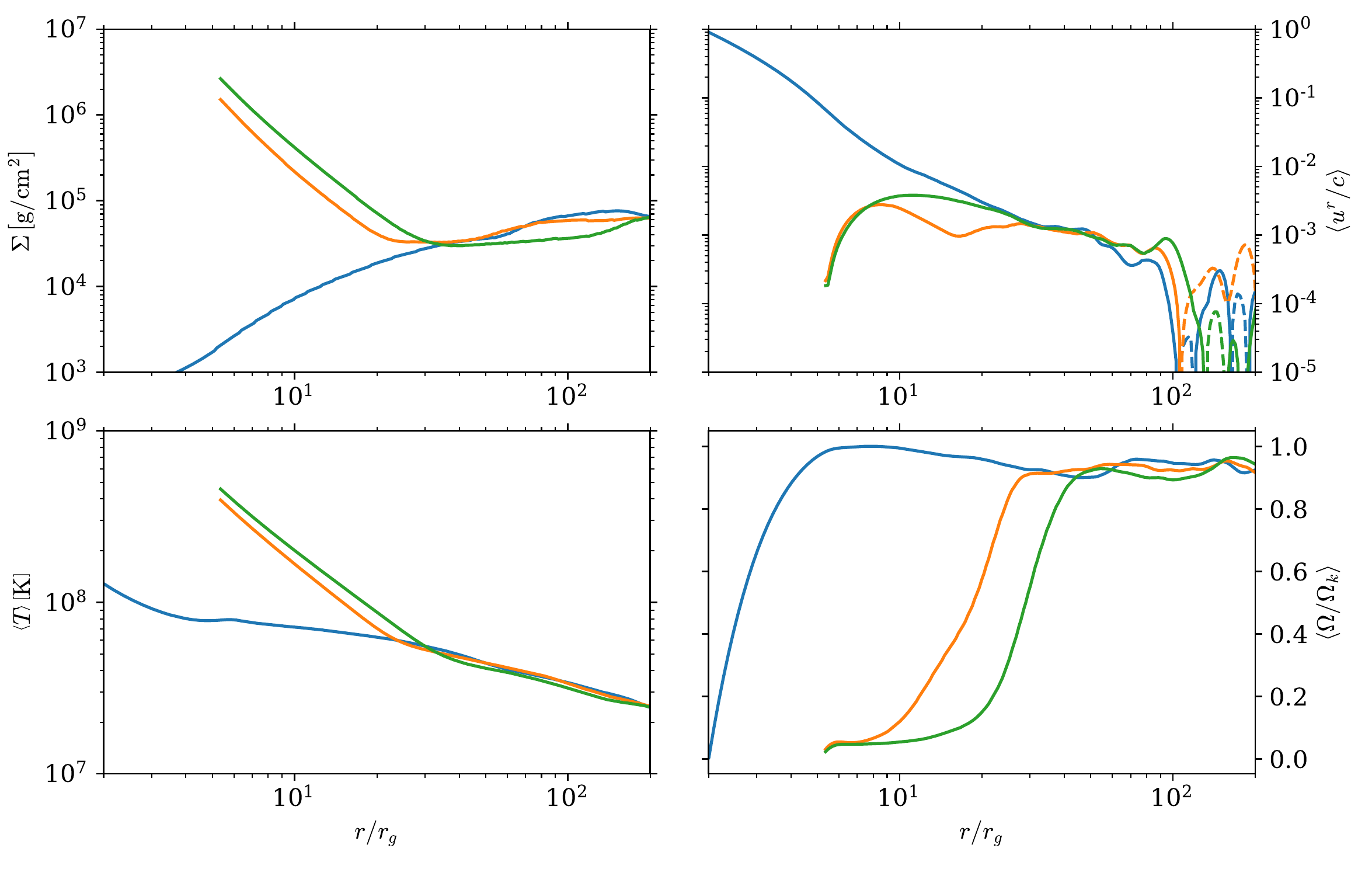}
	\caption{Here we show radial profiles of surface density, $\Sigma$, density-weighted, $\theta$-averaged inflow velocity, the
    		density-weighted, $\theta$-averaged temperature,
    		  and the density-weighted, $\theta$-averaged
    		ratio of angular 
    		momentum to the Keplerian angular momentum. The data is time averaged 
    		over the range, $t_4$ from $40\, 000\, t_g$ to  $80\, 000\, t_g$ 
    		and 
    		$\theta$-averaged over one scaleheight, $h$.
    		Plots are shown for the both \BH{} (blue)
    		and \NS{} (orange). 
    		We also plot in green the same quantities from \NS{} time averaged over the interval
    		$t_5$ from $80\, 000\, t_g$ to $160\, 000\, t_g$.
\label{fig:radial}}
\end{figure*}

In Fig.~\ref{fig:radial} we show the radial profiles of both simulations. 
\DA{\BH{} and \NS{} are shown in blue and orange, respectively, for
the time interval $t_4$. The green lines shows the later evolution of
\NS{} over time interval $t_5$. }
The first quantity shown is the surface density, 
\begin{equation}
\Sigma = \int_{\pi/2-h}^{\pi/2+h} \rho \,r \sin\theta d\theta
\end{equation}
integrated within one scale height. The three remaining plots 
correspond to density weighted, $\theta$ averages, which for a given quantity $f$,
appears as 
\begin{equation}
\left<f\right> = \dfrac{\int_{\pi/2-h}^{\pi/2+h}\rho f \sqrt{-g}d\theta}
{\int_{\pi/2-h}^{\pi/2+h} \rho \sqrt{-g} d\theta} .
\end{equation}
The scale height captures a majority of the mass in 
\BH{} and can be a reasonable approximation for the boundary of the accretion disk. 
It becomes a less useful tool however in the inner edge of \NS{}, where the distribution of
density becomes roughly spherical, and the distinction between disk and transition layer 
requires more information. Nevertheless, this only affects the weighting of the quantities,
which, if they are also roughly spherical, should still give a good measure of the 
radial profile.
We can see that the transition layer surface density follows a power law before settling into
the accretion disk which matches that of the black hole at about radius $r=30\, r_g$. 
The second panel shows $\left<u^r/c\right>$, the density weighted radial
component of the four velocity. The inflow velocity for \NS{} remains nonrelativistic. 
The local maximum at the inner edge
is related to the circularization pattern seen in Fig.~\ref{fig:velavg}.
The third panel shows density weighted temperature,
$\left<T\right>$. The temperature  inside of the atmosphere approaches 
$10^{8.5}$ K, and also follows a power law at the inner edge.
\DA{We also measured the growth rate of the temperature on the surface of the neutron star
and found that it grows with a $T(t)\propto t^{1/4}$ dependence.} 
The last panel shows the density weighted ratio of angular velocity
to the Keplerian angular velocity, $\left< \Omega/\Omega_k\right>$. 
 We can also see the angular velocity
transition towards zero as observed in Fig.~\ref{fig:levo}.  The accretion disk is
mildly sub-Keplerian.

\subsection{Outflow and luminosities}
\label{sec:outflow}
\DA{Accurate measurements of the outflow from black hole and neutron star accretion disks are
important for estimating feedback on the surrounding media of these systems. 
We measure the radiative and 
kinetic luminosities of our accretion flows as well as the mass outflow rates.} We purposely chose a very large 
simulation domain so that we can measure these quantities out to larger radii, however, 
these measurements are meaningless without indication that the simulation has reached
some sort of equilibrium. Typically, this is measured by computing the viscous time
of time averaged data over a particular inteval, and seeing at which radius
 the viscous time is equal the length of that time interval \citep{narayan+12}. 
This roughly corresponds to about $r=80r_g$ for our 
simulations averaged over time interval $t_4$, 
evidence of which can be seen in Fig.~\ref{fig:radial} where the averaged
radial velocity changes sign at about 100 $r_g$, an indication that inflow equillibrium has not been reached
passed this radius. 
This means that we can trust the results of our simulation
inside radius $r=80r_g$. 

Outside $80 r_g$ however, we can still trust the results of
the outflow, as long as it is causally connected to the converged region inside
radius $r=80r_g$. Because the velocity of the outflow is much higher than that of the
inflow, this corresponds to much larger radii where we can reliably measure the outflow.
To quantify this, we measure the density weighted average velocity, 
$\left<u^r\right>$ 
\DA{as a function of radius.} We take the average over the $\theta$ coordinate
only for cells with 
 $Be>0$ to reasonably track outflowing gas. 
 Here, $Be$ is the relativistic Bernoulli number, 
\begin{equation}
Be = - \dfrac{T^t{}_t + R^t{}_t + \rho u^t}{\rho u^t}.
\end{equation}
 In steady state, $Be$ is conserved along streamlines 
and gas with positive $Be$ is
energetic enough to escape to infinity, and so it is a reasonable parameter to define the outflow.
\DA{Seperating the outflow form the disk flow is not precise, so one should not put too much influence
on the Bernoulli parameter. For example, we can see from Fig.~\ref{fig:velavg} that the outflow from
near the neutron star transition layer appears to change Bernoulli parameter. 
We have investigated this further
and found that when looking at the non-time-averaged data
in this particular region, the flow does not appear to be steady, and the
$Be=0$ surface is highly variable. For steady flows (such as the outflow at larger radii, the
Bernoulli parameter should still be appropriate for calculating $r_{cc}$}

\begin{figure}
	\centering
	\includegraphics[width=\columnwidth]{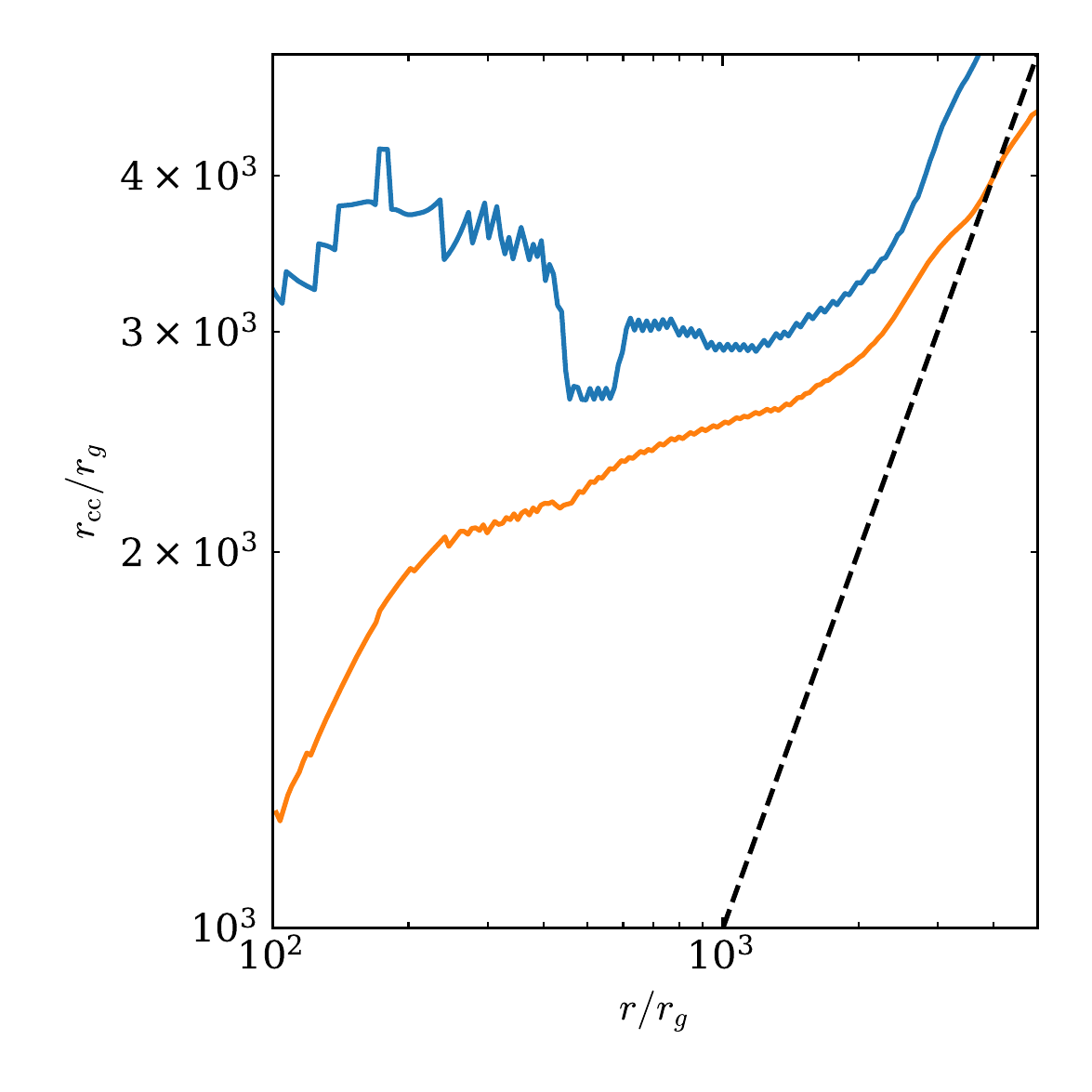}
	\caption{Here we demonstrate the convergence of the outflow by plotting the 
	radius causally connected to the inner region of the disk, 
	$r_\mathrm{cc}$, as a function of the
	radius at which measure the velocity, $r$. 
	For $r_\mathrm{cc}>r$ we can reliably
	measure the outflow of the simulation. For reference, $r=r$ is shown 
	by the black dashed line. As with Fig.~\ref{fig:radial}, blue corresponds
	to \BH{} and orange to \NS{}. \DA{If $r_\mathrm{cc}>r$, then the gas
		had enough time to reach radius $r$.} 
\label{fig:outflowconv}}
\end{figure}

\DA{
We use time averaged data for the time interval, $t_4$,
from $40,000\, t_g$ to  $80,000\, t_g$, and multiply the time and $\theta$-averaged velocity by the duration of the
time interval, $\Delta t = 40,000\, t_g$ to
find the causally connected radius, $r_\mathrm{cc}$, for an outflow of that velocity. This assumes a constant
velocity along the outflowing trajectory, however, since velocity tends to decrease with radius,
the causally connected radius we find is a lower limit.}
Then, as long as the causally connected radius is larger than the radius where we are 
measuring the velocity, we can believe the results of the outflow. A plot of $r_\mathrm{cc}$
is shown in Fig.~\ref{fig:outflowconv}. We can see that the outflow of 
\BH{} is reliable throughout the entire 
\DA{outflow region excluding the region behind the initial
torus shown in Fig.~\ref{fig:velavg}.} For \NS{}, it is
reliable to about $r\approx 4000 r_g$. \DA{It is important to note that this analysis does not 
take into account changes in the nature of the outflow with time, but only that the outflow is causally
connected to the central region of the simulation.}

An additional issue which must be addressed is contamination of the results
by the initial condition. Our initial condition is an equilibrium torus where the total
pressure has been distributed between gas and radiation. The torus is constructed
assuming radiation pressure domination, which implies an adiabatic index, $\gamma=4/3$.
However, the effective adiabatic index depends on the particular mix of gas and radiation
at a particular position, and so does not everywhere equal $4/3$, especially in cooler
parts of the torus. For this reason, the torus is not in perfect hydrostatic 
equilibrium. There is also a certain amount of gas that is blown off of the outer
edges of the torus due to radiation pressure. However, the whole torus was constructed with
$Be<0$, and so due to the lack of dissipation 
or viscosity in the \DA{regions in the outer parts of 
the torus where the MRI has yet to develop}, 
there is a reasonable expectation that $Be$ remains less than zero. Therefore, by measuring
the outflowing quantities over regions with $Be>0$, we are reasonably screening the 
contamination by the initial condition. 
\DA{Additionally, this region overlaps significantly with the area excluded
by the last streamline from the $Be=0$ surface.}
We can see which region this corresponds to by examining the purple dashed 
line in Fig.~\ref{fig:avg}. This region is relatively small compared to the computational
domain of the simulation.
A more robust diagnostic would be for example,
evolving a tracer along with the simulation to track the evolution of gas.
\DA{Then it would be possible to check whether the gas originated from 
the inner, converged regions of the simulations or not.}

\begin{figure*}
	\centering
	
	\includegraphics[width=2\columnwidth]{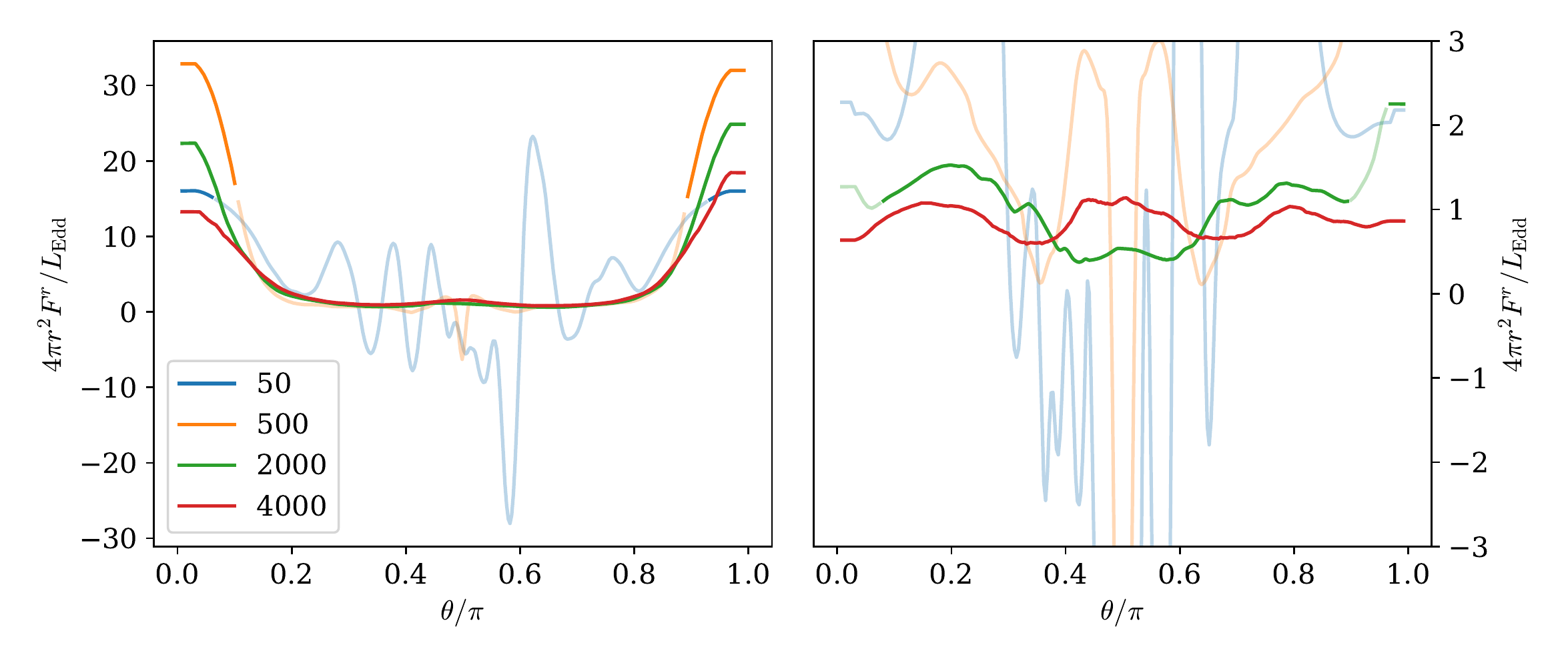} 	
	
	\caption{ Here we show the radial distribution of the flux in units of
		\DA{the apparent isotropic 
			luminosity, $4\pi r^2 F^r/L_\mathrm{Edd}$, as a function of the line of sight}
		for
		\BH{} (left) and \NS{} (right).
		Each line corresponds to the observable flux at a particular radius shown 
		in the legend in units of $r_g$. The faded
		regions correspond to areas where the scattering optical depth is greater
		than one. 
		\label{fig:luminosity}}
\end{figure*}
 
In
Fig.~\ref{fig:luminosity}, we show the angular distribution of radiative flux, $F^r=R^r{}_t$,
\DA{ as a function of $\theta$. The units of the flux are given in the inferred isotropic 
luminosity by multiplying
by $4 \pi r^2$. We plot the flux at four different radii, $50 r_g, 500 r_g, 2000 r_g, 4000 r_g$.}
We have emphasized the regions
where the gas is optically thin, as the radiation in the optically thick regions is 
not expected to reach the observer.  The optically thick regions should not be 
completely ignored however, it is still possible for the outflowing matter to add or
subtract from the radiation field, and so it is difficult to say what the distribution of
radiation will look like at infinity. The \BH{} behaves as expected. The optically thin funnel region produces locally super-Eddington fluxes. We can see that the radiation is highly collimated, even at large radii. 
The intensity of radiation increases with radius at lower radii, and decreases at larger radii,
as the disk and outflowing gas emit and absorb radiation. 
\NS{} is more complicated however. Due to the thick outflows the photosphere is 
pushed nearly to the edge of the simulation domain. Only at very large radii can we 
measure a flux of locally about Eddington. 
Larger accretion rates may be required to reach higher luminosities, 
although it is hard to say whether or not the increase in outflowing gas will cancel
any increase in luminosity. 
Indeed, in the case of black holes, it is well known that at super-Eddington accretion
rates, an increase in the accretion rates corresponds to a decrease in the radiative 
efficiency so that the luminosity scales logarithmically, $L \propto \log(1+\dot{M}/\dot{M}_\mathrm{Edd})$
\citep{ss+73}.
It may even be interesting to try smaller accretion rates in
case the outflow scales differently than the luminosity in the presence of a
reflective boundary condition.

\begin{figure}
	\centering
		 \includegraphics[width=\columnwidth]{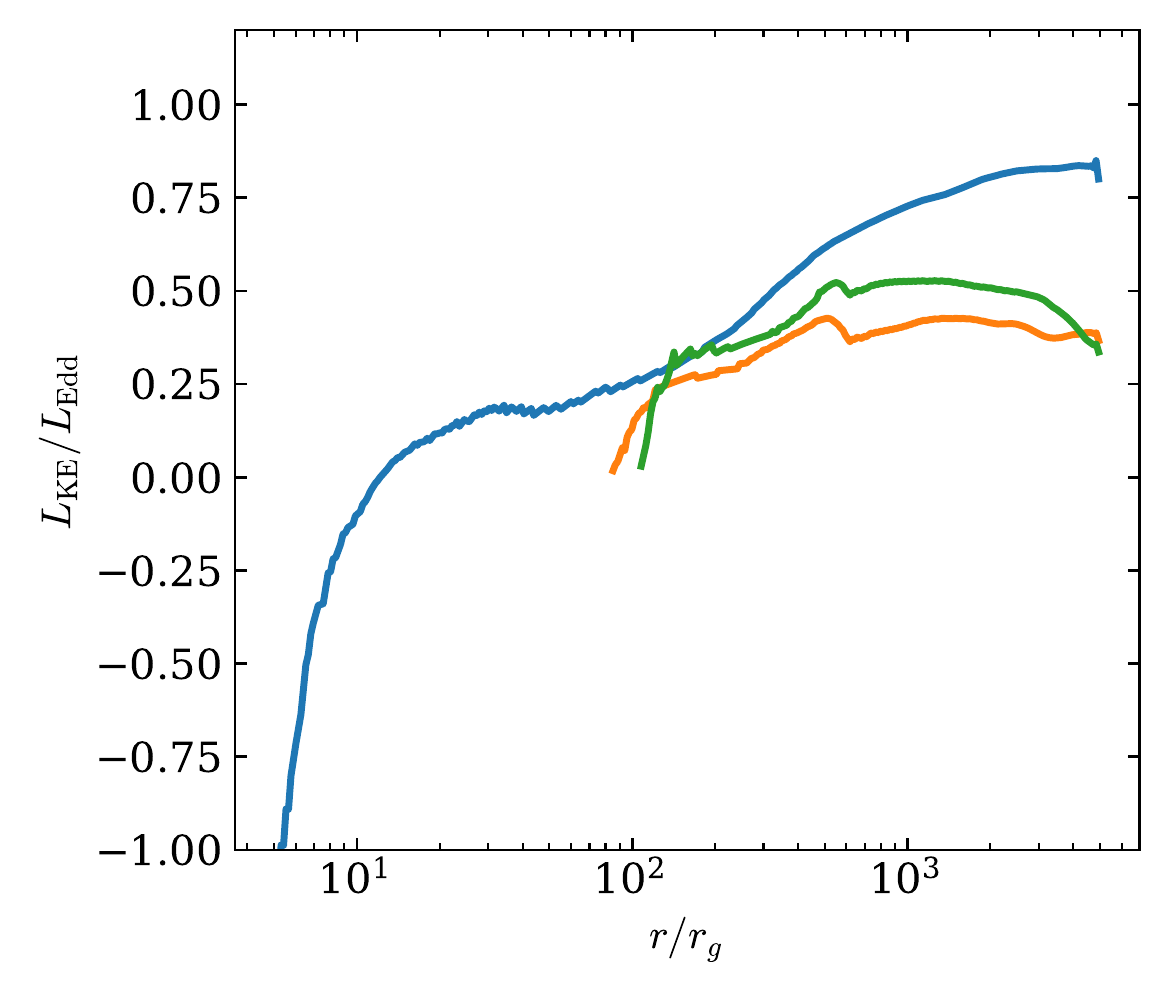} 
	\caption{Here we plot the outflowing luminosity of 
	kinetic energy $L_\mathrm{KE}$, as a 
	function of radius. The kinetic energy flux is integrated over the outflowing
	region defined by positive relativistic Bernouli number, $Be$. \BH{} 
	is shown in blue, and \NS{} is shown in orange. 
		\label{fig:keout}}
\end{figure}

Besides observational properties, the implications of black hole feedback play
an important role in the astrophysics of star formation, star clusters, and 
galaxies. In Fig.~\ref{fig:keout}, we show the luminosity of kinetic energy, 
defined as, 
\begin{equation}
L_\mathrm{KE} =2 \times 2 \pi \int_{0}^{\theta_\mathrm{out}} -(u_t + \sqrt{-g_{tt}})\rho u^r \sqrt{-g}\,\mathrm{d} \theta,
\end{equation}
where $\theta_\mathrm{out}$ is the angle at which $Be=0$. By integrating over positive $Be$,
we are choosing only the gas which is energetic enough to reach infinity.
The extra factor of two reflects the fact that we include the contribution from both sides
of the equatorial plane.
In the non relativistic limit, the integrand approaches $1/2 \rho v^2 v^r$,
where $v^i$ is the three-velocity, $v^2$ is the square of the three-velocity, and $v^r$ is the radial component of
the three-velocity.
\DA{\NS{} shows a large amount of material ejected into the surroundings
of the neutron star environment.  
A large amount of it remains bound} and so only at around $r=100r_g$ is $\theta_\mathrm{out}>0$. 
\NS{} everywhere
has a less energetic outflow than \BH{}, leveling off at $L_\mathrm{KE} \approx 0.4 L_\mathrm{Edd}$. 
For \BH{} the gas is energetic enough to escape
to infinity in the funnel region even at radii below the stagnation radius, however
the velocity vector is directed inward and so the flux of kinetic energy is into
the inner boundary, thus the transition to negative values in the kinetic luminosity.

The total radiative output of the simulation is difficult to measure directly, especially
for optically thick flows. Radiation and gas can still exchange energy even after the gas
has left the simulation domain. We chose to run our simulations with a very large outer 
radius, so that the photosphere is still contained in the simulation. This allows us to
measure the amount of radiative energy outflowing from optically thin and outflowing
regions. Radiation flowing through optically thin regions is expected to escape to 
infinity, and so is an effective lower limit on the total radiative output. We define
the optically thin region by the angle at a particular radius
where the radial scattering optical
depth is equal to one. For each radius, we integrate the radial radiative flux over the 
optically thin region, 
\begin{equation}
L_\mathrm{thin} = 2 \times 2\pi \int_0^{\theta_\mathrm{thin}} -R^r{}_t \sqrt{-g}d\theta.
\end{equation}

A reasonable upper limit of the radiative luminosity, is the
integral of $F^r$ over the region of $Be>0$, $L_\mathrm{out}$, as in the calculation of $L_\mathrm{KE}$.
This gas is energetic enough to reach infinity, and so it 
is possible for the outflowing gas to eventually release its trapped photons. Note that it is
possible for this gas to produce additional photons, but this is not thought to 
contribute significantly to the radiative luminosity, so $L_\mathrm{out}$ is not a strict upper limit.
The outflowing luminosity is then given as, 
\begin{equation}
L_\mathrm{out} = 2 \times 2\pi \int_0^{\theta_\mathrm{out}} 
-R^r{}_t \sqrt{-g} d\theta
\end{equation}

\begin{figure}
	\centering
		 \includegraphics[width=\columnwidth]{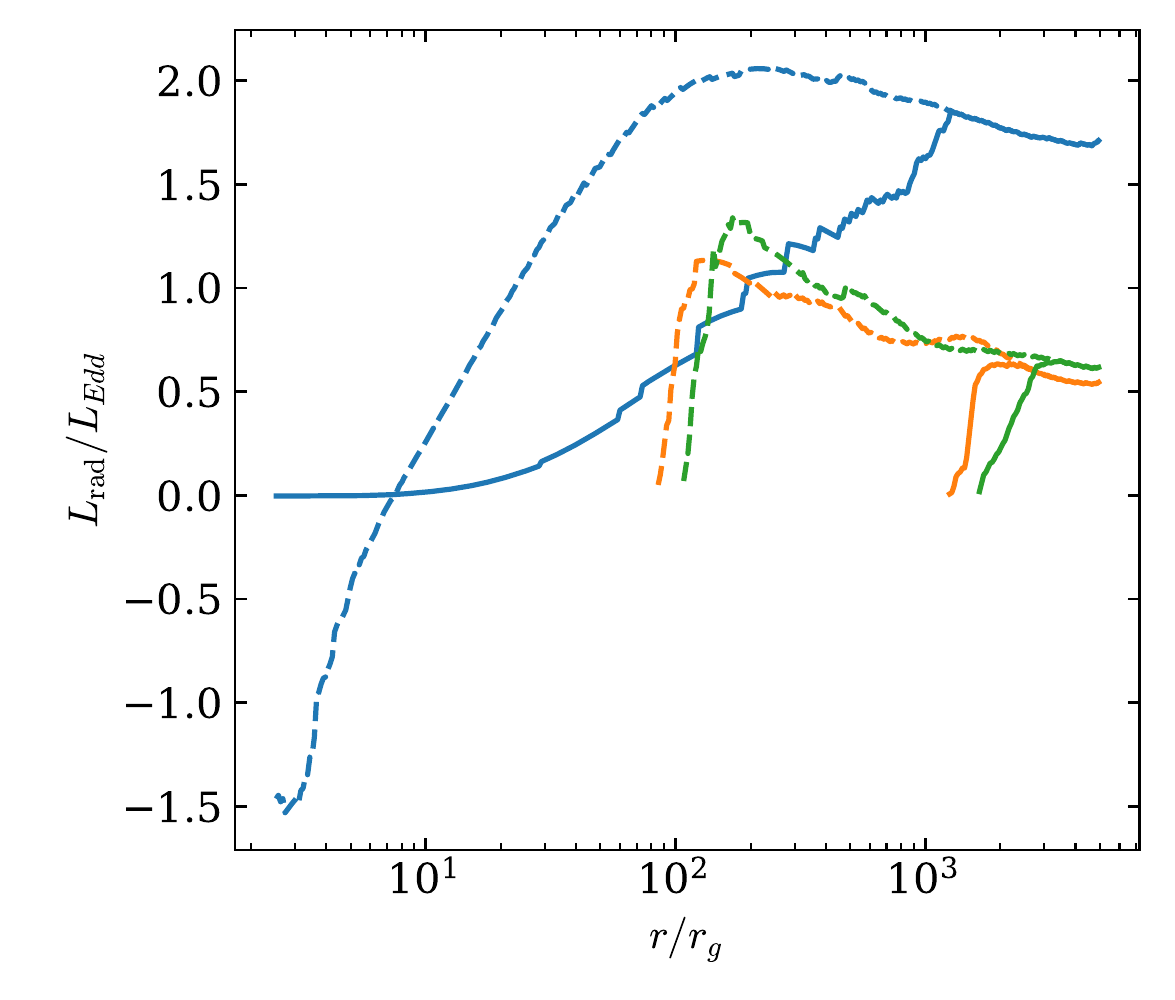} 
	\caption{Radiative luminosities are shown for \BH{} (blue) and 
		\NS{} (orange) averaged over the time period $t_4$. 
		The last time period for \NS{}, $t_5$ is shown in green.		
		The solid lines correspond to the 
		luminosity integrated over angles that are optically thin, 
		$L_\mathrm{thin}$, while the dashed lines
		show the luminosity integrated over the angles with 
		positive $Be$, $L_\mathrm{out}$. 
		\label{fig:lumwr}}
\end{figure}

The outflowing and optically thin radiative luminosities are shown in Fig.~\ref{fig:lumwr}. 
In general \BH{} shows about two to three times as high luminosity  as \NS{}. 
As shown in Figs. \ref{fig:bhframes} \& \ref{fig:nsframes},
\NS{} produces a much larger amount of radiation energy. However, the vast majority 
is trapped in the optically thick outflow and transition layer. 
\DA{We also observe a general decrease in luminosity at radii larger than about $300 r_g$. 
This is because of radiation flowing into the excluded regions behind the initial torus. 
The optical depth of this region is low and so radiation can escape from the outflow into this 
region. It is hard to say how the radiation would behave in this region in reality, it would
depend on the extent and thickness of the accretion disk.}
\DA{We do not seem much change in the radiative luminosity 
	over the $t_5$ time interval.}
%In principle one should not ignore the diffusive flux, which, in the optically
%thick limit can be estimated using the Eddington approximation, 
%\begin{equation}
%F_i = \dfrac{c}{3\rho \kappa}\dfrac{d\widehat{E}}{dx^i}.
%\end{equation} 
%We find however that the diffusive flux account is only relevant for \BH{}. A detailed
%discussion can be found in \citet{sadowski+3d}, on whose simulations we base our own,
%in which they find that the diffusive flux is only important in the transition region
%from optically thick to optically thin, in-between the disk and the funnel region. 
%For \NS{} however, most of this region is already accounted for in $L_\mathrm{out}$. In the
%inner regions of the simulation the diffusive flux account for only a very
%small fraction (0.0001 to 0.001) of the total flux, which is dominated by the advection
%of photons.

\begin{figure}
	\centering
		 \includegraphics[width=\columnwidth]{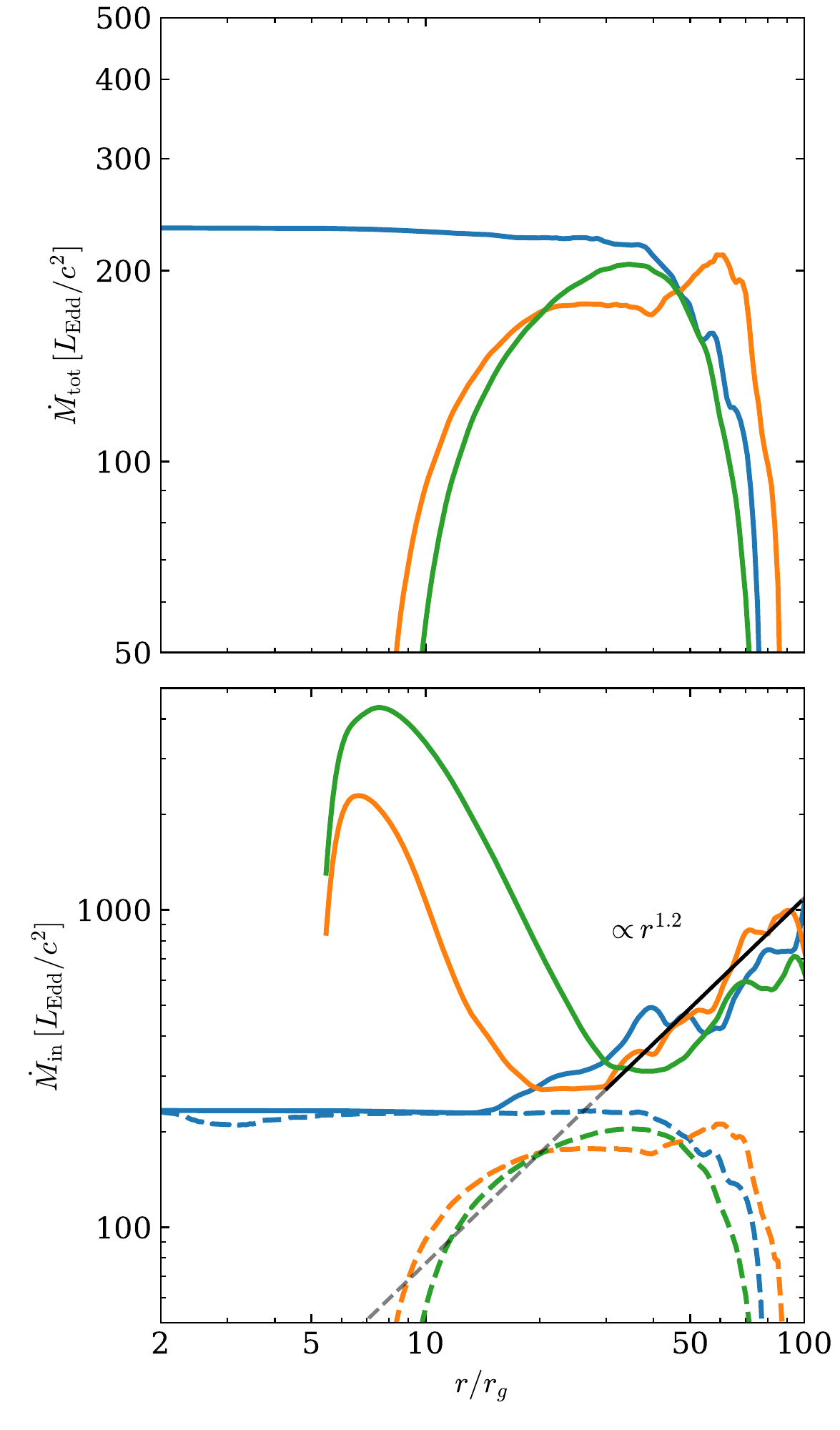} 
	\caption{
	Top: Here we show the total accretion rate,
	$\mdottot$, of \BH{}
	(blue) and \NS{} (orange) computed from data averaged over the time interval, $t_4$. 
	\DA{In green, we show data from \NS{} time averaged over the interval $t_5$. }
	Bottom: Here we plot the mass accretion rate of the inflow, $\mdotin$, again for \BH{}
	and \NS{} from $t_4$ and $t_5$. We include two definitions of the inflow. 
	The first corresponds to the solid lines which
	are defined as the integral over all cells with negative $u^r$, $\dot{M}_{\mathrm{in},u^r}$. 
	The dashed line shows the
	the integral over all cells with negative $Be$, $\dot{M}_{\mathrm{in},Be}$.
	\DA{A power law fit to $\dot{M}_{\mathrm{in},u^r}$ for \NS{} from $t_4$ is shown in black. 
	The solid black line shows
	the radii over which the fit was performed, the dashed part is an extension
	of the power law for reference.} 
		\label{fig:mdot_tot}}
\end{figure}

Super-Eddington accretion flows are known to have strong radiation driven outflows \citep{ohsuga+05,hashizume+15,fiacconi+17}. We
can measure this by examining the accretion rate in a few different ways. First in the top
panel of Fig.~\ref{fig:mdot_tot}, we measure the total (net) mass accretion rate as a function of radius, 
\begin{equation}
\dot{M}_\mathrm{tot} = 2\pi \int_0^\pi -\rho u^r \sqrt{-g}d \theta.
\end{equation}
\BH{}, which is shown in blue, gives a flat accretion rate out to about $r\lesssim 50 r_g$, 
which roughly corresponds to the region where the simulation has achieved 
steady state.
\NS{}, shown in orange, is also nearly flat \DA{for a small region between $20 r_g$ and $50 r_g$.}
At low radii the simulation 
is accumulating mass, which corresponds to a non-flat slope of $\mdottot$. 
In general, $\mdottot$ is lower for \NS{} than for \BH{}. This is not due to a lower accretion rate
in the disk, but due to the fact that the gas that is normally lost through the inner
boundary of \BH{} either accumulates in the transition layer where
it is \DA{recycled into the inner accretion flow}, or is ejected into the outflow, 
the latter of which contributes to a lower value of $\mdottot$.
\DA{We also show $\mdottot$ for \NS{} averaged over the time interval $t_5$. We can see that
the flat region indicating an accretion disk in steady state is no longer present. This is 
due to two effects. The first is the transition layer increasing in radius. The second is due to the fact 
that the outflows are increasing at higher radii and so the total accretion rate over the whole sphere
decreases.}

The second panel of Fig.~\ref{fig:mdot_tot} shows the inflowing accretion rate as measured in two different ways. 
The first is a more naive measurement of the inflow accretion rate, 
$$\dot{M}_{\mathrm{in},u^r} = 2 \pi \int_{u^r<0}  - \rho u^r \sqrt{-g} d\theta,$$
where we simply sum over individually inflowing cells. These are the solid curves in the
second panel, and in general, \BH{} and \NS{} agree with each other outside of the 
neutron star transition layer. 
This indicates that they are accreting at the same accretion rate, the fundamental scale of
accretion, and we can expect the same behavior from their accretion disks.
\DA{The inflowing accretion rate, $\dot{M}_{\mathrm{in},u^r}$, is a good measure to determine the mass accretion
rate in the disk when most or all of the disk turbulence is averaged out, then all of the gas in the disk should
have $u^r<0$, and all of the gas in the outflow should have $u^r>0$. This is not the case for the data we show
except at small radii. 
For flows where the 
turbulence is not fully averaged out, $\dot{M}_{\mathrm{in},u^r}$ is more of an upper limit. A super-Eddington
accretion flow is expected to have an accretion rate that increases linearly with $r$ \citep{ss+73}, and we can
see from the black line, that this dependence is nearly reproduced in \NS{}. 
We obtain $\dot{M}_{\mathrm{in},u^r} \propto r^{1.2}$. }

\DA{We can still see a large amount of gas
flowing through the \NS{} accretion disk for the $t_5$ interval with nearly the same slope as for $t_4$, although
at slightly lower values. This is likely due to the structure of the initial torus, which over very long
periods of time, does not supply the accretion disk with gas at the same rate, slightly dropping with time.
}

The dashed lines correspond to the integral of the total mass flux over the regions with negative
Bernoulli number,
\DA{
$$\dot{M}_{\mathrm{in},Be} = 2 \pi \int_{Be<0}  - \rho u^r \sqrt{-g} d\theta,$$
}
 \DA{Most of the gas at lower radii has negative Bernoulli number, whether or
not it is part of the disk or the outflow. From Fig.~\ref{fig:velavg}, we can see streamlines 
that are initially outflowing with negative $Be$ which turn around and return to the disk midplane, but we
can also see streamlines which seem to change from negative to positive $Be$, and which continue to flow outwards.
It is important to note that $Be$ is only conserved along streamlines for steady flows. Closer to the disk,
the flow is very much non-steady, and so the streamlines do not always reproduce the gas trajectory. For
this reason, one must be cautious when using $Be$ as a diagnotic for outflow. It is necessary to know something
about the flow, and so we put more weight on $Be$ farther from the disk where the flow is more steady. In this 
way $Be$ is more useful as a diagnotistic for outflows at larger radii. We still include plots of $\dot{M}_{\mathrm{in},Be}$ for completeness, although they should mostly match $\mdottot$ at lower radii. 
The departure of  $\dot{M}_{\mathrm{in},Be}$ from $\dot{M}_{\mathrm{in},u^r}$ for \BH{} 
is also evident from the
top-left panel of Fig.~\ref{fig:velavg} where there is gas which has postivie $Be$ but is directed into the
black hole. }
 
\begin{figure}
	\centering
	\includegraphics[width=\columnwidth]{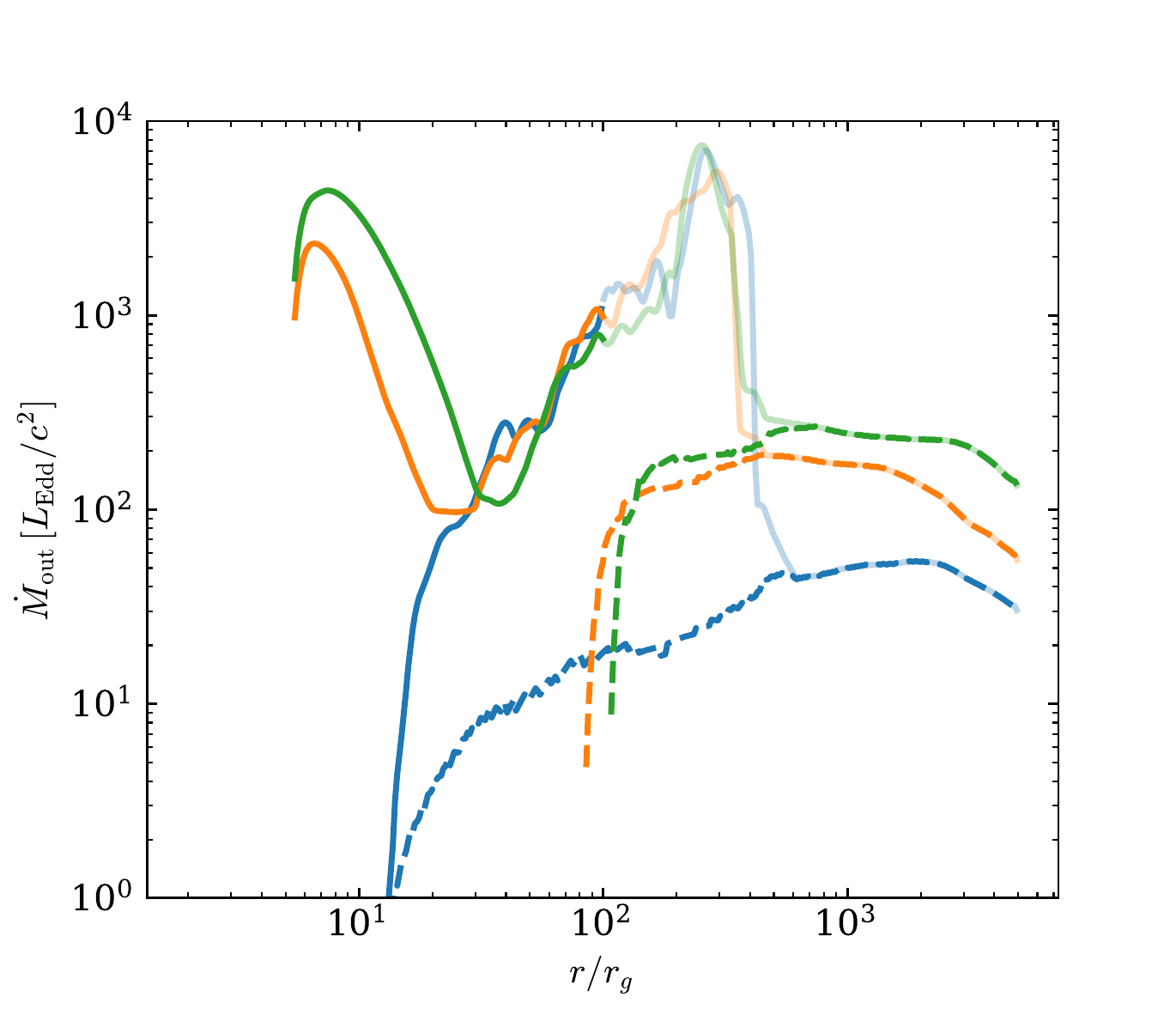} 
	\caption{Here we show the outflowing
		mass accretion rate, $\mdotout$. $\dot{M}_{\mathrm{out},u^r}$ is given by the solid lines
		and $\dot{M}_{\mathrm{out},Be}$ is given by the dashed lines. Blue lines correspond 
		to the \BH{} and orange to \NS{} both using time averaged data from the interval $t_4$. 
		The green lines shows data from \NS{} time averaged over the interval, $t_5$. 
		\label{fig:mdot_out}}
\end{figure}

Fig.~\ref{fig:mdot_out} shows the outflowing accretion rate, measured in the same way as the inflowing
accretion rate, $\dot{M}_{\mathrm{in},u^r}$, but with opposite sign, $\dot{M}_{\mathrm{out},u^r}$. 
\DA{Again we show \BH{} in blue and \NS{} in orange. We also plot in green the mass outflow rate for \NS{} over the 
last interval, $t_5$ from $80\,000 t_g$ to $160\,000 t_g$ which was not run for the \BH{}, 
because it is more computationally demanding.}
\DA{At lower radii, $\dot{M}_{\mathrm{out},u^r}$ more or less follows $\dot{M}_{\mathrm{in},u^r}$, this is because, 
in inflow equilibrium, the difference between them, $\mdottot$, should be constant. 
As the accretion rate drops with a decrease in radius, the outflow rate must also drop
so that no gas can accumulate at any radius, at least for \BH{}. 
We have also demphasized $\dot{M}_{\mathrm{out},u^r}$ at radii larger than $100 r_g$, 
where the simulation has not reached inflow equilibrium.}

For measuring the mass outflow that is expected to reach infinity, the relevant quantity is 
$\dot{M}_{\mathrm{out},Be}$, the total mass outflow rate in the
positive energy region, the region which is separate from the accretion disk. 
\DA{These are the dashed curves in Fig.~\ref{fig:mdot_out}.}
\DA{$\dot{M}_{\mathrm{out},Be}$, is calculated in the same way as, $\dot{M}_{\mathrm{in},Be}$ 
	except for $Be>0$.} In general, \NS{} shows nearly an order of 
magnitude larger outflow rate than \BH{} at its peak when measured over the same time period. 
This is why the outflows are so optically thick. 
The maximum outflowing accretion rate for \NS{} nearly matches \DA{$\mdotin$ at the edge of the transition layer.}
\DA{If the outflow were to converge to this value ($\sim 200 L_\mathrm{Edd}/c^2$)
 at all radii $\gtrsim 1000 r_g$, then this 
would be a strong indication that all the inflowing matter is eventually ejected. 
We do not observe this however,
and moreover we observe that mass is still accumulating at the inner edge.}
While the simulation was run for a long time,
around (160,000$t_g$), the transition layer did not seem to reach a steady state.
\DA{We can see the outflow at large radii become more flat for $t_5$, this also indicates 
that the outflow while apparently increasing with time, may slowly be approaching at 
least a quasi-steady state, and that outflows which are the same magnitude as the accretion rate
in the disk cas be expected for long periods of time.}

\section{Discussion}
We have run two long duration GRRMHD simulations, one of accretion onto a Schwarzschild black hole
and one onto a neutron star (non-magnetized, non-spinning). 
Our black hole simulation, \BH{}, 
is used as a baseline to \DA{compare against} the neutron star simulation for 
two reasons. One, it is very similar to the inflow boundary conditions used in many 
previous simulations. Second, it is a simple, well studied system, without an 
artificial boundary condition, and is directly comparable to other previous simulations
of its type and the \DA{model more closely matches the behavior of the physical object it represents}. 
It is immediately apparent that the simulations
are quite different. The differences in the accretion disk structure are apparent inside
radius $r=30 r_g$, where the flow in \NS{} begins to change into the transition layer. 
It is then inferred that the differences in the outflow are due to the various physical
processes occurring in the inner region of the simulation. We observed from Fig.~\ref{fig:levo}
that the transition layer increases with size as time goes on. 

The simulation was run for \DA{ $120,000\, t_g$}, 
which only corresponds to \DA{one} second of physical time, which is only \DA{one} full pulsation 
for the pulsation periods observed in pulsating ULX systems. These observations were integrated over  kilosecond
time scales.
The point being that even though our simulations are run for a long time 
(when compared to other GRRMHD simulations), they would have to be run much longer in order
to capture the effects of the \DA{varying radiation source} of a rotating magnetized star. 
Such long duration
simulations are likely out of the question for 3D simulations, likely still difficult
for our 2D simulations, and impossible for 2D simulations without any sort of dynamo.
Nonetheless, we are free to speculate.

 \DA{We observe
that the radius outer boundary of the transition layer increases with a time dependence that is well 
described by $t^{0.85}$ power law, and we have no reason not to expect this dependence to continue} 
until either the mass supply is quenched,
the radiation pressure becomes too strong, or the central density and temperature
become large enough that some other cooling mechanism becomes effective, e.g., 
neutrino cooling or \DA{photon cooling through outflows}. \DA{The central temperature
grows with a $t^{1/4}$ power law, and so neutrino cooling, which may become relevant around $10^9\,\mathrm{K}$ may
need to be considered (depending on the central density) after a few (4-8) times the simulation length.}
 We expect the other processes to also take much longer than a few times the current
 simulation time, so it would be interesting to run an even longer duration simulation, 
in order to study the time evolution of the outflow and luminosity as the simulation
evolves. In our case, the length of our simulation was chosen so that the outflow would
be causally connected to the inflow equilibrium region of the inner simulation to 
radii past the photosphere. \DA{A non-physical limitation would be the transition layer growing faster
than inflow equillibrium. A potential solution would be to run a black hole like simulation
for long time and then to restart the simulation with an inner boundary.}

Our main conclusion from our work is that the presence of a hard, nonrotating surface alone is 
likely not sufficient to produce enough radiation that can escape to infinity to explain
the luminosities observed in ULX sources. 
The large amount of ejected
gas not only obscures the innermost emitting regions, but also decollimates the radiation,
further reducing the observed luminosity. Thus it is vital that we consider the effects of
magnetic fields. This was done by \citet{takahashi+17}, who found a bolometric luminosity
to be about an order of magnitude larger than ours, the simulation is short however, 
$t_\mathrm{stop} = 15,000 t_g$, so more work would be required to study the effective of
large amount of gas being accreted onto the magnetic field. 

It would be interesting to study the case of accretion onto a non-magnetized neutron star
with a different initial condition for the disk magnetic field. Our initial condition contains
magnetic field loops of alternating polarity, and reconnection is allowed on a numerical 
level. Thus, no significant amount of magnetic flux accumulates on the inner edge of the 
simulation, indeed the magnetic pressure is several order of magnitude lower than
the radiation pressure in the transition layer. 
 Running a simulation with only one magnetic field loop, which normally leads 
to accretion in the magnetically arrested disk (MAD)
 state in black hole simulations \citep{mckinney+15,narayan+17}, would likely also lead to
 a buildup of magnetic flux in the transition layer which could be dynamically important.
 
 \DA{
 One interesting application of our study could be an extension to supersoft sources (SSS) \citep{long+81,heuvel+92}
 and ultraluminous supersoft sources (ULS) \citep{urquhart+16}. ULSs are characterized by peak temperatures
 of around $10^6$ K, with bolometric luminosities at a few times $10^{39}$ erg s$^{-1}$ and photospheres 
 at radii of around $10^4$ km. SSSs are classified similarly with lower temperatures 
 ($\sim 10^5$ K, $\sim 10^6$ K) and luminosities ($\sim 10^{36}$ erg s$^{-1}$, $\sim 10^{38}$ erg s$^{-1}$). 
 The effective temperature of \NS{} can be computed by measuring the radiative energy density around
 the photosphere which is on average around a few times $10^6$ K. The radius of the photosphere is around
 5000 km. These features are indicative of ULS sources, although the bolometric luminosity is about an order
 of magnitude too low, closer to the higher end for SSSs. While SSSs can be well explained by 
 nuclear burning on the surface of a white dwarf \citep{heuvel+92}, ULSs, like ULXs have been harder to model. 
 It would be interesting in future work to do post-processing of \NS{} to generate spectra
 to see if they match any of the more specific characteristics of ULSs or SSSs. 
 }

\subsection{Comparision to \citet{takahashi+17b}}
After this work was completed we became aware of an important
publication by \citet{takahashi+17b} (TMO18), 
who obtained similar results to ours by running a shorter simulation (7000 $t_g$),
demonstrating the feasibility of super-Eddington accretion onto a non-magnetized neutron star.
It is vital that we make a comparison to the work by TMO18, 
as the simulations they have run should be directly comparable to ours. TMO18 have performed two 
simulations, one onto a black hole, and one onto a reflective inner boundary as in our work. 
Both codes are GRRMHD solvers with $M_1$ closure. The code used is 
the same as used by the authors in their previous work. 
Both our simulations and those of TM018 show a power law behavior at the inner edge
of the simulation.
TMO18 report a significant increase in the rate of ejected mass over their black hole simulation.
Indeed we also infer a roughly ten times increase in the mass outflow rate for \NS{} 
over \BH{}. 
The authors also
report an increase in the radius of the photosphere over the black hole case. We find a
much larger increase in the photosphere however, reaching out to a couple thousand
$r_g$. This is likely due to the length of our simulations, which are run for more than ten
times the length of the simulations in TMO18, and so we believe we are able to make a 
more accurate statement about the mass outflow to larger radii, and about the radiative 
luminosity, which TMO18 does not address. 

\DA{One interesting difference is found in the strong change in the angular velocity
in the transition layer, seen in our simulations. TMO18 show the centripetal force, which can 
be used as a proxy for $\Omega$ (indeed, in the Newtonian case it is proportional to $\Omega^2$). TMO18 show 
that the centripetal force transitions from super-Keplerian to about half Keplerian. This contrasts against
our simulations where $\Omega$ transitions to practically zero at the inner edge of the simulation, even at 
early times. This difference could be due to differences in the inner boundary condition for $u^\phi$.
When we average our data over the same time interval as is described in TMO18 (3000-5000 $t_g$)
and we examine $\Omega/\Omega_K$ we see that it transitions 
from unity at about 10 $r_g$ to 0.3 at the stellar surface ($5r_g$).
This corresponds to the same range of radii over which the centripetal force of TMO108
transitions to about 0.6 times the Keplerian value which would
indicate an $\Omega/\Omega_K\approx0.8$, more than twice the
angular velocity ratio of our simulation.
This indicates that we have a stronger torque
at the inner edge of our simulation, and so it is probably easier for 
gas to accumulate in the transition layer. It will be beneficial in the future
to study angular momentum transfer through the inner boundary
for different boundary conditions.}

\subsection{Caveats}
It is important to stress that our simulations are a first attempt at measuring the effects
of a hard surface in the context of a global, radiative, MRI-driven, super-Eddington accretion
flow. More physics is planned in future implementations. Indeed there are a number of issues
that must be addressed before we may consider these results truly robust.

First, our simulations are performed in 2D axisymmetry. \cite{sadowski+3d} has shown 
that non axisymmetric effects can lead to a lower luminosity accretion flow 
around a black hole. Since the luminosities we measured were on the low end
of what is expected for ULX objects, we cannot be sure if a three dimensional simulation
would be bright enough. It is possible that 
\DA{this difference would not occur in the presence of hard surface}, but a fully 
three dimensional simulation would be required to confirm. 

Second, we have neglected the effects of rotation. Accretion onto a rapidly rotating star
will release less energy than accretion onto a stationary star. The gas will also be 
less bound however. For a one second rotational period, the effects of rotation at the
neutron star surface should be negligible; however, the acceleration of gravity scales as
$r^{-2}$ while centripetal acceleration increases with $r$. At larger radii, this may become
relevant if the accreted atmosphere rotates nearly uniformly. 

\section{Summary}
We have found that a reflective, non-rotating boundary at the inner edge of an accretion disk 
simulation has a significant effect on the behavior and structure of the
inner disk as well
as its emission and outflows. 
\DA{We observe large amounts of gas accumulating on the inner boundary
	in a transition layer, where the angular velocity transitions from its Keplerian
	value to near zero.}
\DA{We also found lower rates in the outflow of kinetic energy over the 
	black hole case. However, we did measure larger mass outflow rates, 
	affecting the release of radiation to the observer.}
In fact, the hard surface
of a non-magnetized neutron star leads to lower
radiative luminosities in super-Eddington flows relative to black holes, 
and with radiation decollimated to the point where they are not likely to explain
even the lowest luminosity ULXs, \DA{although they may be 
	applicable to ULSs or SSSs}.
This work is a first step in a larger plan to study 
accretion processes around a neutron star. Further work includes studying different
accretion rates, including rotation, and eventually the addition of stellar magnetic fields.

\section*{Acknowledgements}
The first author thanks Ramesh Narayan, Maciek Wielgus, and Jean-Pierre Lasota for 
advice and illuminating conversations. Research supported in part by the Polish NCN grants 2013/08/A/ST9/00795
and 2017/27/N/ST9/00992. Computations in this work were performed on the Prometheus machine, part of the PLGrid infrastructure.

%%%%%%%%%%%%%%%%%%%%%%%%%%%%%%%%%%%%%%%%%%%%%%%%%%



\bibliographystyle{mnras}
\bibliography{mybib} % if your bibtex file is called example.bib

%%%%%%%%%%%%%%%%%%%%%%%%%%%%%%%%%%%%%%%%%%%%%%%%%%

%%%%%%%%%%%%%%%%% APPENDICES %%%%%%%%%%%%%%%%%%%%%

%\appendix


%%%%%%%%%%%%%%%%%%%%%%%%%%%%%%%%%%%%%%%%%%%%%%%%%%

% Don't change these lines
\bsp	% typesetting comment
\label{lastpage}
\end{document}